\documentclass[prx,amsmath,amssymb,aps,twocolumn,balance]{revtex4-1} 

\usepackage{latexsym}
\usepackage{amsmath, amsthm, amssymb} 
\usepackage{txfonts}
\usepackage{natbib}
\usepackage{epsfig}
\usepackage{graphicx}
\usepackage{amsmath}
\usepackage{amsfonts}
\usepackage{amssymb}
\usepackage{cancel}
\usepackage{graphicx}
\usepackage{ulem}
\usepackage{wrapfig}
\usepackage{epsfig}
\usepackage{graphicx}
\usepackage{amsmath}
\usepackage{amsfonts}
\usepackage{amssymb}
\usepackage{graphicx}
\usepackage{wrapfig}
\usepackage{textcomp}
\usepackage{filecontents}
\usepackage{pgf,pgfarrows,pgfnodes,pgfautomata,pgfheaps,pgfshade}
\usepackage[colorlinks=true, linkcolor=blue]{hyperref}
\usepackage{lipsum}
\usepackage{balance}

\newcommand{\be}{\begin{equation}}
\newcommand{\ee}{\end{equation}}
\newcommand{\bml}{\begin{multline}}
\newcommand{\eml}{\end{multline}}

\begin{document}

\title{
Coulomb blockade of  chiral   Majorana and complex fermions far from equilibrium}

\author{Dmitriy S. Shapiro$^{1,2}$} \email{dmitrii.shapiro@kit.edu} 
\author{Alexander D. Mirlin$^{1,3}$} \author{Alexander Shnirman$^{1,3}$}
\affiliation{$^1$\mbox{Institute for Quantum Materials and Technologies (IQMT), Karlsruhe Institute of Technology, 76021 Karlsruhe, Germany}}
 \affiliation{$^2$National University of Science and Technology MISiS,   119049 Moscow, Russia}  
\affiliation{$^3$Institut f\"ur Theorie der Kondensierten Materie, Karlsruhe Institute of Technology, 76128 Karlsruhe, Germany} 

\begin{abstract}

 We study charge  transport    in 
 a single-electron transistor   implemented as an interferometer such that the Coulomb blockaded middle island contains a circular   chiral Majorana or   Dirac  edge mode.
  We concentrate on the regime of small conductance and provide an asymptotic solution in the limit of     high transport  voltage exceeding the charging energy. 
 The solution is achieved using      an instanton-like technique.
 The distinctions between Majorana and Dirac cases appears when the tunnel junctions are unequal. The main difference is in the offset current at high voltages which can be higher up to 50\% in Majorana case.
 It is   caused by an additional particle-hole  symmetry of the distribution function  in the Majorana case.  There is also an eye-catching distinction between the  oscillations patterns of the current as a function of the gate charge. We conjecture this distinction  survives   at  lower transport voltages  as well.

\end{abstract}

\maketitle

\section{Introduction}

   The effect of Coulomb blockade in a single-electron transistor (SET)~\cite{kulik1975kinetic,Averin:1986aa,ingold1992charge,PhysRevB.54.5428,Kouwenhoven1997,ALEINER2002309,PhysRevLett.82.1245}, a device where  Fermi-liquid leads are mediated by a quantum dot, plays an essential role in condensed matter physics, mesoscopics     and open quantum systems.
The Coulomb spectroscopy and transport through a quantum dot are  sensitive to the precise nature of the non-equilibrium steady state, the mechanisms of relaxation, electronic interactions  and topological order~\cite{ALTLAND20062566,sedlmayr2006tunnelling,PhysRevLett.102.026805,PhysRevB.82.155317,PhysRevB.84.165440,PhysRevLett.114.176806,Albrecht:2016aa,PhysRevB.93.155428,PhysRevB.95.075425,PhysRevLett.122.016801,PhysRevLett.124.196801,PhysRevResearch.2.023221,DOTDAEV2021127736}.
The ``orthodox" theory of    Coulomb blockade is based on rate equations  formulated in the   basis of different charged states in the island~\cite{kulik1975kinetic,Averin:1986aa,ingold1992charge,Kouwenhoven1997}. Such states are well defined for almost isolated quantum dots which have weak tunnel coupling to the leads, i.e, with a small dimensionless conductance, $g{\ll}1$. In this theory, the  distribution function  in the   island   is not affected by the the coupling to contacts, i.e.,  the internal  relaxation in the island is assumed to prevail  on the characteristic tunneling timescale. In  multi-channel limit with $g{\gg}1$, when the Coulomb blockade is weak and the charge is ill defined, the description via the dissipative Ambegaokar-Eckern-Sch\"on (AES) action~\cite{PhysRevLett.48.1745,PhysRevB.30.6419}  for the phase  becomes more convenient~\cite{ALTLAND20062566}.  In equilibrium,  the   saddle points of  the Matsubara AES action are known as Korshunov instantons~\cite{korshunov1987coherent}. These  instantons  allow one to take into account charge discreteness   and obtain the residual, exponentially small  gate charge oscillations of the conductance. 
If the relaxation is weak then    the distribution function in the island is a non-Fermi one at finite voltages. It causes a non-equilibrium steady state  that can not be captured by the ``orthodox" theory or the imaginary time technique, and consequently the real time  Keldysh formalism~\cite{keldysh1965diagram,Kamenev}  is required.  Important recent achievements include  the  theoretical analysis of strongly non-equilibrium transport    using the AES action~\cite{PhysRevLett.102.026805}, and the generalization of Korshunov instantons to the real-time Keldysh formalism~\cite{PhysRevB.93.155428} at $g{\gg}1$. 
In particular,  the results of Ref.~\cite{PhysRevLett.102.026805} lead directly to the conclusion that the Coulomb blockade is lifted at  transport  voltages lower than in the ``orthodox" theory due to  the non-equilibrium distribution function in the island. 

 In this work, we study the  strongly non-equilibrium regime   of high voltages     that  exceed significantly the   charging energy  of an  island, and we assume the strong Coulomb blockade, i.e., the dimensionless conductance is small, $g{\ll} 1$. 
 At low voltages one thus expects a strong suppression of the charge transport. 
At higher voltages the almost Ohmic behavior is accompanied by the offset (deficit) current and   residual gate charge oscillations.
Here, we were able to describe this high voltage regime asymptotically exact. 
 Instead of using the kinetic equations, which    is challenging  due to a large number of  relevant charge states, we employ a path integral technique and succeed solving the problem by finding a dominant path, an alternative kind of  instanton, for the phase variable.

 \begin{figure}[h!]
	\center{\includegraphics[width=0.95\linewidth]{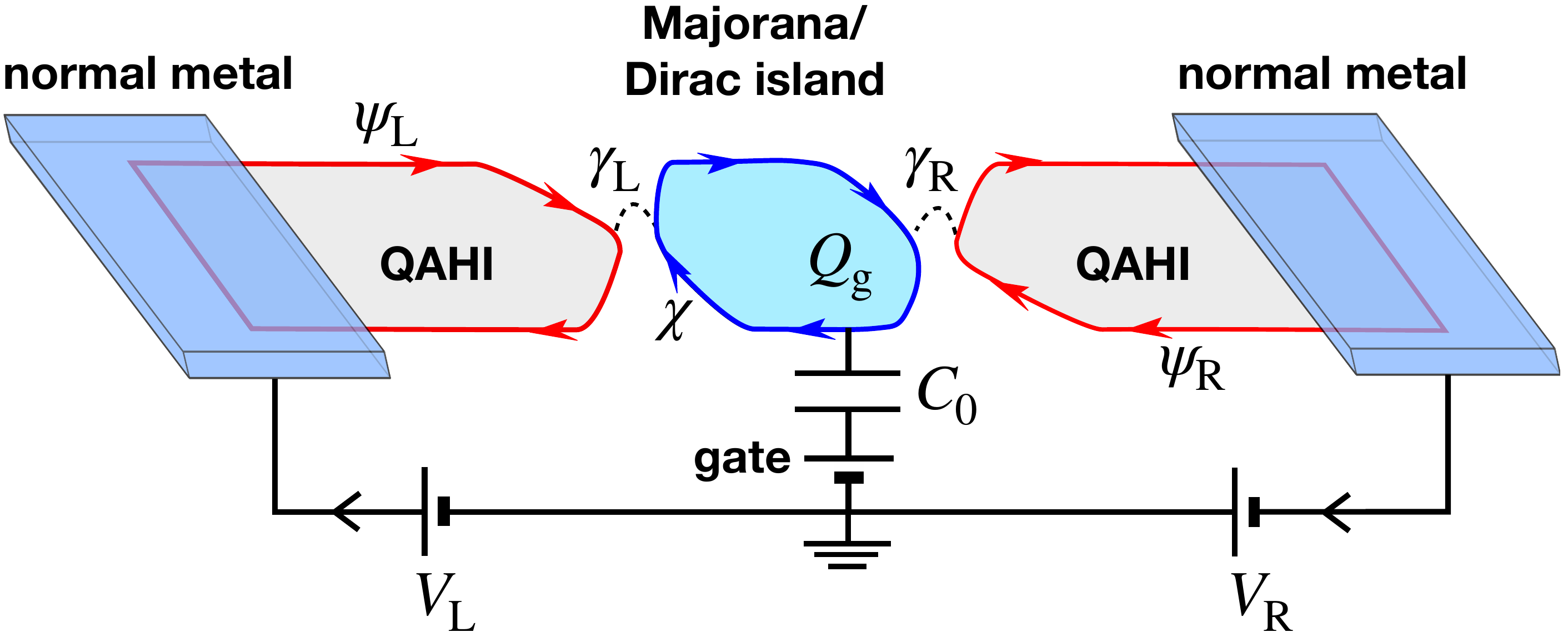}}
	\caption{Sketch of single-electron transistor realized as chiral Majorana or Dirac interferometer. Normal metal leads (Ohmic contacts) cover spinless channels with  chiral Dirac fermions,  $\psi_{\rm L,R}$, which  are  the edges of a quantum anomalous Hall insulator (QAHI) film. Gate voltage controls the offset charge $Q_{\rm g}$ in the   island of  capacitance $C_0$. The island is in the topological superconducting  or normal state. The chiral   mode  $\chi$ hosts   Majorana or   Dirac    fermions. Tunnel amplitudes are $\gamma_{\rm L,R}$, and bias voltages are symmetric, $V_{\rm L(R)}{=}{\pm}V/2$, the positive direction of the current is marked by black arrows.
}
	\label{setup}
\end{figure}

We apply our solution for calculations of the non-equilibrium tunneling density of states {\color{black}(TDoS)} and current-voltage relations using the formalism developed by Meir and Wingreen~\cite{PhysRevLett.68.2512}.
The devices under consideration are chiral interferometers  implemented in hybrid structures with superconductors,  topological insulators or quantum anomalous Hall insulators (QAHI)~\cite{FuKanePRL2008,HeScience2017,shen2018spectroscopic}  (see Fig.~\ref{setup}).      They  
have two Fermi-liquid leads biased by the voltages $\pm V/2$, and tunnel coupled to the central island. The  island   hosts a single conducting channel which  is either a real Majorana fermion edge mode {\color{black}(the case when the island is in a   proximity induced topological superconducting phase)} or a complex Dirac one {\color{black} (the case of a normal island in a   quantum-Hall topological insulator state)}.  There is an electrostatic gate that induces an offset charge in the island, $Q_{\rm g}$. The charging energy of the island is $ E_{\rm c} =e^2/(2C_0)$ with $e$ is electron charge and $C_0$ is the total capacitance of the island.
  The chiral fermions propagate   with a velocity $v$ along the edge channels.  In this case, the  Thouless energy, $E_{\rm Th}=\frac{2\pi \hbar v}{L}$ ($\hbar$ is Plank constant),    is nothing but  level spacing in the ring of the perimeter $L$. (Hereafter we set   $e{=}\hbar{=}1$.) We assume  metallic spectrum of the  edge modes which means that  $E_{\rm Th}$ is sufficiently small.  The voltages  are limited from above by the energy scale  $\Delta_0${\color{black}---the absolute value of the superconducting order parameter induced in the topological part of the island---}above which   other    conducting channels or 2D scattering states become relevant. Ultimately, we work under  the following assumption, 
\be\Delta_0 >eV\gg \{ E_{\rm Th},  E_{\rm c}  \} \ . \label{condition}
\ee  
 Moreover, we assume  
 no relaxation to phonons, no electron-electron  scattering and zero  temperature. 
 The only relaxation mechanism is due to the  tunneling to the   leads.  
 In this regime the single-particle distribution function is expected to develop a multi-step structure which will play a substantial role below.

\section{Model}
\subsection{Keldysh action for the Majorana device}
The microscopic description of the charge transport  is provided by the  Keldysh  generating functional
\begin{equation}
\mathcal{Z}[\eta]{=}   
\!\! \int\! D[\Psi  ]   D[\chi ]   D[\varphi] e^{iS[\Psi,\chi ,\varphi  , \eta ]}\ . \label{Z0}
\end{equation}
The first path  integral is  taken over complex fermions{\color{black}, $\Psi{=}\{\Psi_{\rm L},\Psi_{\rm R} \}$,} collected in Nambu spinors  $\Psi_{\rm L}{=}\{ \psi_{{\rm L}, k}, \bar \psi_{{\rm R}, - k} \}$  and $ \Psi_{\rm R}{=}\{ \psi_{{\rm R}, k}, \bar\psi_{{\rm R}, -k} \}$ where $\psi_{{\rm L}, k}$ ($\psi_{{\rm R}, k}$) are   Grassmann fields in left (right) leads; these are chiral states of momenta $k\in [-\infty, \infty]$. The second  variable $\chi(x)$  is Majorana edge mode in the island being real Grassmann field  defined on a ring with a coordinate  $x\in [0,L]$. The third one,  $\varphi$, is {\color{black}the phase of the} superconducting order parameter in the island.
$\mathcal{Z}$ depends on a pair of counting fields $\eta_L$ and $\eta_R$ (source variables) collected in  $\eta=\{\eta_{\rm L},\eta_{\rm R}\}$. They   generate the charges  $Q_{l}$ that   flow from  the left ($l=\rm L$) or  right  ($l=\rm R$) lead into the island during  the measurement interval $t{\in}[0;t_0]$:
\be
Q_l= \left. i \frac{\partial \mathcal{Z}[\eta]}{\partial \eta_l}  \right|_{\eta\to 0}  \ . \label{Q}
\ee
The total action is \begin{equation}
 S= S_{\rm c}+ S_{\rm L}+ S_{\rm R}+ S_{\rm M} +  S_{\rm L}^{\rm (tun)}+ S_{\rm R}^{\rm (tun)} \ .
\end{equation}
In the Keldysh technique, the   physical  time  $t\in [-\infty, \infty]$ gets doubled, $t_\pm$, with the index $\pm$ denoting the  upward and backward parts of Keldysh contour  $\mathcal{C}$. 
Then, the Keldysh rotation  to classical and quantum components of the boson field, $\varphi_{\rm cl}$ and $\varphi_{\rm q}$, is performed:
$\varphi(t_\pm)=\varphi_{\rm cl}(t)\pm \varphi_{\rm q}(t)/2$.

A coherent dynamics  of $\varphi$  is governed by the action
$S_{\rm c}= \int \mathcal{L}_{\rm c}[\varphi] dt $  
 where the Lagrangian is
\be
\mathcal{L}_{\rm c}=  
   \frac{ \dot \varphi_{\rm q} \dot\varphi_{\rm cl} }{ 8 E_{\rm c}  } 
- \frac{1}{2}    \dot \varphi_{\rm q} Q_{\rm g}     \ .
\ee
Complex fermion dynamics  is described  by Keldysh actions, $S_l{=}\int_{\mathcal{C}}dt  ( \int \frac{dk}{2\pi} \bar\psi_{l,k} i \partial_t \psi_{l,k}  - H_l)$,  where $H_l{=} \hbar v\int \frac{dk}{2\pi} k \bar \psi_{ l, k} \psi_{ l,k} $ is a Hamiltonian of   chiral fermions and  the lead index  is $l{=}{\rm L,R}$. 
The action in $\pm$-basis reads $S_l{=}\!\sum\limits_{\sigma,\sigma'}\!\iint \frac{d\omega dk}{(2\pi)^2}  
\bar\psi_{l, \sigma} [G^{-1}_{l,\omega,k}]_{\sigma,\sigma'} \psi_{l, \sigma'} $ where {\color{black}the inverted fermion Green function is}
\be
{G}_{{\rm L(R)},\omega,k
	}^{-1} = (\omega{-}vk)\sigma^z{+}i \varepsilon( (\sigma^0+\sigma^x) f_{{\rm L(R)}, \omega} -i   \sigma^y), \label{Gl}
\ee
 $\sigma^\alpha$ are the Pauli matrices, $\varepsilon{>}0$ is an infinitesimal broadening constant.   We keep the $\pm$-basis for fermions while perform the Keldysh rotation to the classical and classical components for the phase, $\varphi$. The functions $f_{l} = 1{-}2n_{l}$ are determined by zero temperature 
Fermi distribution functions in the  leads,  $n_{\rm L(R),\omega} {=}\frac{1}{2} {-}\frac{1}{2} {\rm sign}(\hbar\omega{\mp }eV/2)$. We assume here that the symmetric voltage bias  $V_{\rm L(R)}{=}{\pm}V/2$ is applied. The energies $\hbar\omega$ of  Dirac electrons in the leads are  counted from a  chemical potential at zero bias.

{\color{black}
We apply a uniform gauge transformation (see Appendix~\ref{AGT}) for the complex fermions $c(\mathbf{r},t)$ in the island, $c(\mathbf{r},t) {\to} e^{-i\mu t -i\frac{1}{2} \varphi(t)} c(\mathbf{r},t) $. After the transformation,     the floating phase $\varphi(t)$ and the yet unknown chemical potential of the $s$-wave superconducting island, $\mu$,   are  eliminated from the island's action; they   appear instead in the tunneling action below. One arrives at the stationary Bogolyubov-de Gennes Hamiltonian, which at low energies (below $\Delta_0$, see Eq.~(\ref{condition})) reduces to the following effective action for   chiral edge Majorana modes:} 
\begin{equation}
   S_{\rm M} = \! \frac{1}{2}
\sum\limits_{\sigma,\sigma'} \int\limits_0^L  \! dx \! \int \! dt   \chi_{\sigma}  
 (i\partial_t{+}i v\partial_x)\sigma^z_{\sigma,\sigma'}
  \chi_{\sigma'}     \ . \label{SM}
\end{equation}
{\color{black}
Here,} the Lagrangian is diagonal in the Keldysh space, i.e., there is no internal  broadening/dissipation. 
 In this case, this kernel is not invertible which means  fully unitary dynamics. An introduction of an infinitesimal broadening with some distribution functions is not necessary as it will anyway be overridden by the coupling to the leads. 
The latter is   described by the tunnel action {\color{black}for the low-energy modes, where the above mentioned gauge transformation and the rotation to the Majorana basis is performed,}
\begin{multline}
S_{  l}^{\rm (tun)}[\varphi,\eta] = \gamma_{  l}  \sum\limits_{\sigma,\sigma'}  \int \! \!  dt  \chi_{l,\sigma}(x_l)  
\!\left(   U[\varphi,\eta_l ]_{\sigma,\sigma'} \psi_{l,\sigma'}^{(0)}  \right. -\\ -\left.      U^+[\varphi,\eta_l ]_{\sigma,\sigma'}  \bar \psi_{l,\sigma'}^{(0)}\right) .
\label{St}
\end{multline}
Here, Majorana  field $\chi(x)$ couples to  the  local fermions in the leads  $\psi^{(0)}_l{=}\int\frac{dk}{2\pi}\psi_{l,k}$ at two points,  $x{=}x_{ \rm L, R}$.
The  tunnel amplitudes $\gamma_l$ are chosen real.    
{\color{black}The matrix \be 
U[\varphi,\eta_l ]= e^{-i\mu t -i\frac{1}{2} \varphi_{\rm cl}(t)}e^{-i\frac{\varphi_{\rm q}(t)+2\eta_l z(t) }{4}\sigma^z}  \label{U}
\ee
captures the gauge transformation   mentioned above and the counting field.}
The yet unknown chemical potential of the island $\mu$ will be found after solving an appropriate kinetic equation. This transformation also means that all energies in the island are counted from   $\mu$.
The charge counting variable $\eta_l$ is an amplitude of the auxiliary quantum field, $\frac{\eta_l z(t)}{2}\sigma^z$. It generates the transmitted      charge $Q_l[\varphi]$ which is  a classical observable. 
Further,   $z(t){=}1$ for $t\in [0,t_0] $ and $z(t){=}0$ otherwise. It switches the charge counting on and off    at $t{=}0$ and $t{=}t_0$, respectively.

\subsection{Non-equilibrium  effective theory for the phase}
After the integration over the  complex fermions, $\Psi$, and then over the     Majorana ones, $\chi$, the generating functional (\ref{Z0}) becomes
\begin{equation}
\mathcal{Z}[\eta]
= \int  D[\varphi] e^{iS_{\rm c}[\varphi]+\frac{1}{2} {\rm tr }\ln  ((i\partial_t{+}i v\partial_x)\sigma^z
-  \Sigma[\varphi , \eta]  )} \ . \label{trln}
\end{equation}
Here, $\Sigma$ is the  self-energy  for Majorana fermions. It reads
 \begin{equation}
 \!\!\!\!\!\Sigma[\varphi , \!\eta ]_{x,t,t'} {=} \sum\limits_{l={\rm L,R}}   \gamma_{l}^2 \left(\mathcal{G}'_l[\varphi ]_{t,t'}{-}\big[\mathcal{G}'_l[\varphi ]_{t'\!,t}\big]^T\right) \delta(x{-}x_{ l})  , \!\!\! \! \label{Sigma}
\end{equation}
where $
\mathcal{G}'_l[\varphi ]_{t,t'} {=}  U_l^+(t) \sigma^z   {\mathcal G}_{l}(t{-}t') \sigma^z   U_l(t') 
$   is the boson-dressed  Green function of the lead.
 The self-energy $\Sigma$ is singular at the points where the tunnel contacts are located.
 The presence of two contributions to   the self-energy ($\mathcal{G}'$ and $\big[\mathcal{G}'\big]^T$) reflects the Majorana nature of the island excitations.  In the time domain  the local   Green functions of the leads read $   {\mathcal G}_{l}(t)  =\int \frac{d\omega}{2\pi}  {\mathcal G}_{l;\omega} e^{-i\omega t}  $ where 
\begin{equation}
  {\mathcal G}_{l;\omega}{\equiv}\!\!\int\frac{dk}{2\pi}   G_{l;\omega,k}   
 {=}\frac{i}{4\pi v} ((\sigma^x-\sigma^0)f_{l;\omega} -i\sigma^y) . \label{G-lead}
\end{equation}

To analyze the phase dynamics,  we develop here an   expansion scheme for the      logarithm in  (\ref{trln}). A naive expansion in the small tunneling amplitude $\gamma_l$ would force us  to introduce an infinitesimal broadening in the island with an arbitrary distribution function. However, in such an approach a physical distribution function, dictated by the leads, would emerge only after the infinite summation of higher order contributions. Instead, we extract from $  \Sigma[\varphi , \eta]$ a part with a constant in time classical phase $\varphi_0$, $\Sigma[\varphi , \eta]{=}\Sigma[\varphi_0 , 0]{+}(\Sigma[\varphi , \eta]{-}\Sigma[\varphi_0 , 0])$. We transfer  the extracted part, $\Sigma[\varphi_0 , 0]$, into the zeroth order propagator~\cite{PhysRevB.95.075425}, 
\be \mathbf{G}_{\rm 0}^{-1}{=}(i\partial_t{+}i v\partial_x)\sigma^z{-}  \Sigma[\varphi_0 , 0] ,  \label{G0-1}
\ee which we can invert exactly, and perform an expansion in $\delta\Sigma[\varphi , \eta]{=}\Sigma[\varphi , \eta]{-}\Sigma[\varphi_0 , 0]$. Due to the  gauge invariance of the problem,   $\varphi_0$ does not appear in the final results.

We   expand  the logarithm in (\ref{trln}) up to the first order in $\delta\Sigma$. After that, $\delta\Sigma$ itself is expanded up to the linear order in  $\eta$ since we are interested in the current only.  Omitting  a constant term, we obtain the following result for the logarithm in (\ref{trln}):
 \begin{multline}
\!\!\!\frac{1}{2} {\rm tr }\ln \left((i\partial_t{+}i v\partial_x)\sigma^z{-}  \Sigma[\varphi , \eta ] \right) 
\! \approx \\ i S_{\rm AES}[\varphi] {-}i\eta_{\rm L} Q_{\rm L} [\varphi]{-}i\eta_{\rm R}Q_{\rm R}[\varphi].  \!\! \label{AES}
 \end{multline}
The first term in (\ref{AES}) is the  dissipative  AES action~\cite{PhysRevLett.48.1745,PhysRevB.30.6419},
\be
S_{\rm AES}[\varphi]  =  i \frac{1}{2} {\rm tr } \Big[\mathbf{G}_{\rm 0 }\delta\Sigma[\varphi , \eta=0] \Big] , \label{Sdiss}
\ee 
 and the second and third terms contain the charges $ Q_{l}[\varphi]$  (cf. Eq.~(\ref{Q})) calculated for a {\color{black}certain}  path,  $\varphi_{\rm c}(t)$ and  $\varphi_{\rm q}(t)$. These are given by
 \be
 Q_{l}[\varphi]=\frac{i}{2}\lim\limits_{\eta\to 0}\partial_{\eta_l} {\rm tr } \Big[\mathbf{G}_{{\rm 0}} \delta\Sigma[\varphi, \eta] \Big]. \label{Qtr}
 \ee
 
   We will denote the frequencies related to the island by $\epsilon$ keeping $\omega $ for the leads. This helps us to remember that the energies on the island are counted from the chemical potential.    Since $\Sigma $ is singular in coordinate representation (cf. Eq.~\ref{Sigma}), one needs to know     the Green function $ \mathbf{G}_{{\rm 0},\epsilon}$ at coincident coordinates, $x{\to}x_l$ and $x'{\to}x_l$.   Comparing the tunneling self-energy and the ballistic propagator, we conclude that the weak tunneling limit corresponds to the condition $\gamma_l{\ll} v$. 
  This  limit    is fully equivalent to the condition of  small broadening of  levels compared to the distance between them, $\frac{\sqrt{\gamma_{\rm L}^2{+}\gamma_{\rm R }^2}}{L}{\ll}E_{\rm Th}$.  In this regime, the Green function reads {\color{black}(see Appendix~\ref{GLOC})}:
 \be
\mathbf{G}_{{\rm 0},\epsilon}=- i \frac{E_{\rm Th}}{2v}  \sum\limits_n\delta(\epsilon-\epsilon_n)
((\sigma^0 - \sigma^x ) f_{{\rm M}, \epsilon_n} + i \sigma^y)  , \label{G-loc-2}
  \ee
   which involves   the four-step function
    \be
f_{{\rm M},\epsilon}=\frac{ \gamma_{\rm L}^2(f_{{\rm L}, \mu{+}\epsilon}{-}f_{{\rm L}, \mu{-}\epsilon}){+}\gamma_{\rm R}^2(f_{{\rm R}, \mu+\epsilon}-f_{{\rm R}, \mu{-}\epsilon})}{2   (\gamma_{\rm L}^2+\gamma_{\rm R}^2 )}  \label{fM}
\ee
describing the   non-Fermi distribution of Majorana fermions. It has  the symmetry, $f_{{\rm M},\epsilon}{=}{-}f_{{\rm M},-\epsilon}$,  which is preserved for  arbitrary $\gamma_{\rm L}$ and $\gamma_{\rm R}$. 

The function $\mathbf{G}_{{\rm 0},\epsilon}$ is singular at $\epsilon{=}\epsilon_n$ where  $\epsilon_n{=}(n{+} n_\nu/2) E_{\rm Th}$ are the  energy levels of the island ($n{\in} \mathbb{Z}$ and  $n_\nu$ is the number of vortices).  To calculate the chemical potential we neglect the fluctuations of phase and assume the constant trajectory $\varphi_{\rm cl}{=}\varphi_0$ and $\varphi_{\rm q}{=}0$. Then we employ the   the charge conservation constraint, $Q_{\rm L}[\varphi_0] {+} Q_{\rm R}[\varphi_0] {=} 0$, for $t_0{\to }\infty$ (cf. Eq.~(\ref{Qtr})) and obtain   
\be \mu= \frac{\gamma_{\rm L}^2-\gamma_{\rm R}^2}{2(\gamma_{\rm L}^2+\gamma_{\rm R}^2)}V \ .
\ee
 Unlike in  the   Dirac case, in which the distribution function is a two-step one governed exclusively by the voltages in the leads, in the current Majorana case there are more steps and the chemical potential is important.

\section{Formalism and Analytical Results}
\subsection{Dissipative action}
We evaluate the AES action of Eq.~(\ref{Sdiss}), and obtain
\be
S_{\rm \! AES}   = 
\iint\!\! dtdt' \! 
\begin{bmatrix}
u^*_{{\rm cl}}& u^*_{{\rm q}}
\end{bmatrix}_t
\begin{bmatrix}
0 & \alpha^A \\ \\
\alpha^R  & i\alpha^K_{{\rm M}}
\end{bmatrix}_{t-t'}
\begin{bmatrix}
u_{{\rm cl}}\\
\\ u_{{\rm q}} 
\end{bmatrix}_{t'}
 \label{Kd}
\ee
where $u_{\rm cl}\!=\!e^{-\frac{i}{2} \varphi_{\rm cl}  } \cos \frac{\varphi_{\rm q}}{4} $ and $u_{\rm q}\!=\!-ie^{-\frac{i}{2} \varphi_{\rm cl} } \sin \frac{\varphi_{\rm q}}{4}$ are classical and quantum parts of gauge exponents. In the quasi-classical limit, the   off-diagonal terms in (\ref{Kd}) determine the dissipative dynamics of the phase. They are given by the retarded (advanced) functions, $\alpha^{R(A)}(t){=}\int\frac{d\epsilon}{2\pi}e^{-i\epsilon t}\alpha^{R(A)}_\epsilon$, 
with the  spectra
\be
{\rm Im}\alpha_\epsilon^{R(A)} = {\pm} \frac{\Gamma}{4} \sum\limits_n \big( 
f_{{\rm D}, E_n+\epsilon }
{-} f_{{\rm M},{\it E_n}} \big) \label{alphaR}
\ee
 (see Appendix \ref{ATR}).  The   diagonal  term is the Keldysh one responsible for  non-equilibrium fluctuations. 
Its spectrum reads 
\be
\alpha_{{\rm M},\epsilon}^K = \frac{\Gamma}{2} \!\sum\limits_n \! \big(1{-}  f_{ {\rm M}, {\it E_n}} 
f_{ {\rm D}, {\it E_n+\epsilon}} \big) \ . \label{alphaK}
\ee
Here, dimensionless $\Gamma{=}\frac{\gamma_{\rm L}^2+\gamma_{\rm R}^2}{2\pi^2 v^2}$ is the coupling strength of the phase to  the   dissipative environment.  
It is  the   probability for a Majorana excitation  to leave the island after encircling it once. In the tunneling limit, we have $\Gamma{\ll}1$. Also, the two-step  distribution of Dirac fermions has been introduced, 
\be
 f_{{\rm D},{\epsilon}} =  \frac{ \gamma_{\rm L}^2 f_{{\rm L}, \epsilon+\mu} +\gamma_{\rm R}^2 f_{{\rm R}, \epsilon+\mu}}{\gamma_{\rm L}^2+\gamma_{\rm R}^2} .
\ee 
In   the metallic limit, i.e.,   small level spacing, we replace the sum over $n$ by the  integral, $E_{\rm Th}\sum_n\!\to\! \int dE$, and  we obtain  the   Ohmic spectrum, ${\rm Im}\alpha_\epsilon^R =  \frac{\Gamma}{2}\epsilon$. 
 The Keldysh kernel reads \begin{multline}
\!\!\!\!\! \alpha^K_{\rm M,\epsilon}{=}
\frac{\Gamma}{2(\gamma_{\rm L}^2+\gamma_{\rm R}^2)^2} \Big[\gamma_{\rm L}^2\gamma_{\rm R}^2(|\epsilon{+}V|+|\epsilon{-}V|+2|\epsilon{+}2\mu|)+\\
 + \gamma_{\rm L}^4(|\epsilon|{+}|\epsilon{+}2\mu{-}V|)+ \gamma_{\rm R}^4(|\epsilon|{+}|\epsilon{+}2\mu{+}V|)\Big],
 \end{multline}
As will be shown below, phase trajectories that contribute to the path integral have typical time scales ${\sim}  E^{-1}_{\rm c} $. {\color{black}One can show that for $\epsilon{\lesssim}E_{\rm c}$ the kernel $\alpha_{\epsilon}^K$ can be replaced by its zero frequency value, $\alpha_{\rm M,\epsilon=0}^K{=}\frac{\Gamma}{2} \xi_{\rm M}  |V|$, provided 
\be
V\gg E_{\rm c} {\rm max} \{ h, h^{-1}\} \ .
\ee
} Here,   
the asymmetry parameter  is defined as  \be
h{=}\frac{\gamma_{\rm R}^2}{\gamma_{\rm L}^2} \ . 
\ee 
The parameter $\xi_{\rm M}$ is given by $\xi_{\rm M}{=}p_h q_h$ where \be p_h {=}\frac{4h}{(1{+}h)^2}\ee and \be q_h{=}1{+}\frac{|1 - h| }{2(1+h)} \  \label{qh}.\ee
Neglecting the frequency dependence of the kernel $\alpha_{\rm M,\epsilon}^K$ at high voltages is similar to the common approximation  for a classical noise at high temperatures. 
In this case, the kernel in (\ref{Kd})   becomes local, and the  AES action assumes the form, $S_{\rm AES}{=}\int \mathcal{L}_{\rm d}[\varphi] dt$, where the  effective dissipative  Lagrangian $\mathcal{L}_{\rm d}[\varphi]$ reads
 \be
\!\!\!\mathcal{L}_{\rm d} [\varphi]   =
     \frac{\Gamma}{2}\left(i\xi_{\rm M} |V| \sin^2 \!\frac{\varphi_{\rm q}}{4} - \frac{i \dot\varphi_{\rm q}}{4}  \cos\! \frac{\varphi_{\rm q}}{2}- \frac{\dot\varphi_{\rm cl}}{2} \sin \!\frac{\varphi_{\rm q} }{2} \right) . \!\!\label{Ld}
\ee 
{\color{black}At low voltages, $V\lesssim E_{\rm c} {\rm max} \{ h, h^{-1}\}$, the AES kernel  becomes  non-local in time and our approach ceases to be accurate.}

\subsection{Formula for the current. {\color{black}Tunneling} density of states}
Due to the non-zero $E_{\rm c}$ no charge accumulation can occur on the island in the long time limit. Therefore, at $t_0{\to}\infty$ we expect $\langle Q_{\rm L} \rangle {=} {-}  \langle Q_{\rm R}\rangle $.
Then we are allowed to consider the following symmetrized form for the current,
$
I{=} t_0^{-1}\frac{\gamma_{\rm R}^2 \langle Q_{\rm L} \rangle- \gamma_{\rm L}^2 \langle Q_{\rm R}\rangle }{\gamma_{\rm L}^2+\gamma_{\rm R}^2} 
$.
Here, the average is taken over all trajectories,    $\langle O\rangle = \int D[\varphi] e^{ i \int (\mathcal{L}_{\rm c}[\varphi] +\mathcal{L}_{\rm d}[\varphi])dt} 
O[\varphi]$. 
After some algebra 
with Eq.~(\ref{Qtr}), we arrive at the formula for the current
\be
I_{\rm M}\!= g \int \nu_{\rm M,\omega} (n_{\rm L,\omega}{-} n_{\rm R,\omega} ) d\omega   \label{IM}
\ee
where the dimensionless conductance is $g{=}\frac{\gamma_{\rm L}^2\gamma_{\rm R}^2}{4\pi^2 v^2 (\gamma_{\rm L}^2+\gamma_{\rm R}^2)}$.
As shown by Meir and Wingreen~\cite{PhysRevLett.68.2512}, the non-equilibrium state of the island is hidden in the   normalized   {\color{black}TDoS}   $\nu_{\rm M,\omega}$. In the metallic limit,  we obtain  
\be
\nu_{\rm M,\omega}{=}1 {-} \frac{1 }{4\pi } \iint   e^{i(\omega-\epsilon-\mu) \tau}  \mathcal{D}(\tau)   f_{{\rm M}, \epsilon} d \epsilon d\tau . \label{nuM}
\ee
The non-trivial contribution  in (\ref{nuM}) is provided by 
\be 
\mathcal{D} (\tau){=}  P^<(\tau)-P^>(\tau).  \label{D}
\ee
where    $P^\lessgtr(\tau){=}\langle e^{-\frac{i}{2} \varphi(\tau_{\pm})+\frac{i}{2}\varphi(0_{\mp})}  \rangle$ are the bosonic propagators. 
 For brevity, the phase variables are written in $\pm$-basis.  The propagators obey   the following symmetry, $P^\lessgtr(-\tau){=}(P^\lessgtr(\tau))^*$.
 
  We note that the phase propagators $P^{\gtrless}$ can be written as the averages, $P^{<}(\tau){=}\langle   b(0)   b^\dagger(\tau)\rangle$ and $P^{>}(\tau){=}\langle   b^\dagger (\tau)  b(0) \rangle$, with  the Heisenberg bosonic operators $b {=}e^{\frac{i}{2}\varphi}$ and $ b^\dagger{=}e^{-\frac{i}{2}\varphi}$. These   are the  ladder  operators of the complex  bosonic mode   acting in a space of different charge states. Hence, the Fourier transformed function $\mathcal{D}_\omega{=}\int\mathcal{D} (\tau)e^{i\omega \tau}d\tau$ describes the non-equilibrium excitations spectrum in the capacitor.

\subsection{\color{black}SET with the Dirac island}
So far, we  have considered the Majorana edge mode in the island. Here, we provide the analogous calculations for a  device with a  normal island that hosts a Dirac edge mode.  The action  $S_{\rm D}{=}\! \sum\limits_{\sigma,\sigma'}\!\int\limits_0^Ldx \int \frac{d\omega}{2\pi} \bar\chi_{\sigma}(i\partial_t+iv\partial_x)\sigma_{\sigma,\sigma'}^z\chi_{\sigma'}$ is expressed now in  terms of a complex field $\chi\neq\bar \chi$. There are following  distinctions from the Majorana case. First, the non-equilibrium distribution function has the  well-known double step structure, $f_{{\rm D},{\epsilon}}$, 
which  is not particle-hole symmetric, i.e. $f_{{\rm D},{\epsilon}}{ \neq} {-} f_{{\rm D},{-\epsilon}}$, except the  limit of fully symmetric setup, $\gamma_{\rm L}{=}\gamma_{\rm R}$.
Second, in the   formula for the current,
\be 
 I_{\rm D}= g\int \nu_{\rm D,\omega } (n_{\rm L,\omega}{-} n_{\rm R,\omega} )  d\omega ,   \label{ID}
\ee
we have 
$
\nu_{\rm D,\omega } = 1 - \frac{1 }{4\pi } \iint   e^{i(\omega-\epsilon-\mu) \tau} \mathcal{D}(\tau)  f_{{\rm D}, \epsilon} d \epsilon d\tau 
$.
Note that the chemical potential does not influence the result in the Dirac case. Indeed, introducing $\tilde f_{\rm D,\epsilon}{=}f_{\rm D,\epsilon-\mu}$, i.e., counting the energy from zero, we see that the distribution function $\tilde f_{\rm D,\epsilon}$ does not depend on $\mu$.
The third distinction is that the prefactor $\xi_{\rm M} {=}p_hq_h$   in (\ref{Ld}) is replaced by 
\be
\xi_{\rm D} {=}p_h \ .
\ee
 It follows from  the Keldysh kernel in the Dirac  case, which reads
\be
\alpha^K_{{\rm D},\epsilon}=\Gamma\frac{(\gamma_{\rm L}^4+\gamma_{\rm R}^4) |\omega| + \gamma_{\rm L}^2\gamma_{\rm R}^2 (|\omega-V| + |\omega+V|)}{(\gamma_{\rm L}^2 + \gamma_{\rm R}^2)^2}.
\ee
{\color{black} Similar to the Majorana case, the frequency dependence of this kernel can be neglected and our approach is accurate if $V{\gg}  E_{\rm c} {\rm max}  \{h,h^{-1} \}$.}

\subsection{Path integration and the instanton. {\color{black}Boson propagator}}
\label{Path-int}
 In this {\color{black}Subsec.~\ref{Path-int}} we consider the cases of Majorana and Dirac in parallel, thus, we omit the indices ``M" and ``D" in $\xi$.
 Calculation of boson exponents  in (\ref{D}) is based   on the following representation of the path integral,
\be\langle e^{-\frac{i}{2} \varphi(\tau_\pm)+\frac{i}{2}\varphi(0_\mp)}  \rangle \!=\!\! \int\!\! D[\varphi] e^{i \mathcal{S}_\pm[\varphi_{\rm q}]+ \int\dot \varphi_{\rm cl} \mathcal{A}[\varphi_{\rm q}] dt}.  \label{exp}
\ee
Here, we have introduced 
\be
\!\!\!\mathcal{S}_\pm{=}{\mp}\frac{\varphi_{\rm q}(0){+}\varphi_{\rm q}(\tau)}{4} {+}\frac{1}{2} \!\int\!\Big(   i \xi  \Gamma   |V| \sin^2 \frac{\varphi_{\rm q}}{4} +  \dot\varphi_{\rm q} Q_{\rm g}\Big) dt ,\!\!
\ee
which does not  contain classical components of the phase. We have also introduced 
\be
\mathcal{A}=\frac{\dot\varphi_{\rm q}(t)}{8 E_{\rm c} }- \frac{\Gamma}{4} \sin \frac{\varphi_{\rm q} (t)}{2}+\frac{\theta(t{-}\tau)-\theta(t)}{2},
\ee
which couples linearly to $\varphi_{\rm cl}$. 
The linearity of the action    in  (\ref{exp}) with respect to   $\varphi_{\rm cl}$ plays the    central role in our solution.  We remind that the exclusively linear dependence on $\varphi_{\rm cl}$ is based on two approximations:  the high  voltage  and     the Ohmic spectrum of the island. 
The linearity in $\dot\varphi_{\rm cl}$  allows one to integrate   this field out and obtain a functional  delta-distribution,  $\int\! D[\varphi_{\rm cl}] e^{i \!\int\! \dot\varphi_{\rm cl} \mathcal{A}[\varphi_{\rm q}]dt }{ =} \delta (\mathcal{A}[\varphi_{\rm q}])$. Therefore, the remaining path  integral over $\varphi_{\rm q}$ is restricted by a manifold of trajectories    satisfying $\mathcal{A}[\varphi_{\rm q}]{=}0$ with the boundary condition $\varphi_{\rm q}(-\infty){=}0$. Analysing $\mathcal{S}_\pm$ we now restrict the allowed trajectories. We note that  $\mathcal{S}_\pm$ has an imaginary part $\sim\! i\int\! \sin^2\frac{\varphi_{\rm q}}{4} dt$. It provides a selection rule for the trajectories: $\exp(i\mathcal{S}_\pm){\neq}0$ 
only if   $\varphi_{\rm q}(+\infty)=4\pi n$, $n\in \mathbb{Z}$. 
We find that only a single solution of the first order differential equation $\mathcal{A}[\varphi_{\rm q}]{=}0$  satisfies this selection rule, $\varphi_{\rm q}(t){=}\Phi_\tau(t)$. 
Therefore, the result of  the  path  integration {\color{black}for the boson propagators} reads   
\be
P^\gtrless(\tau)= e^{i  \mathcal{S}_\pm[\Phi_\tau(t)] } \ . 
\ee

\begin{figure}[h!]
	\center{\includegraphics[width=0.95\linewidth]{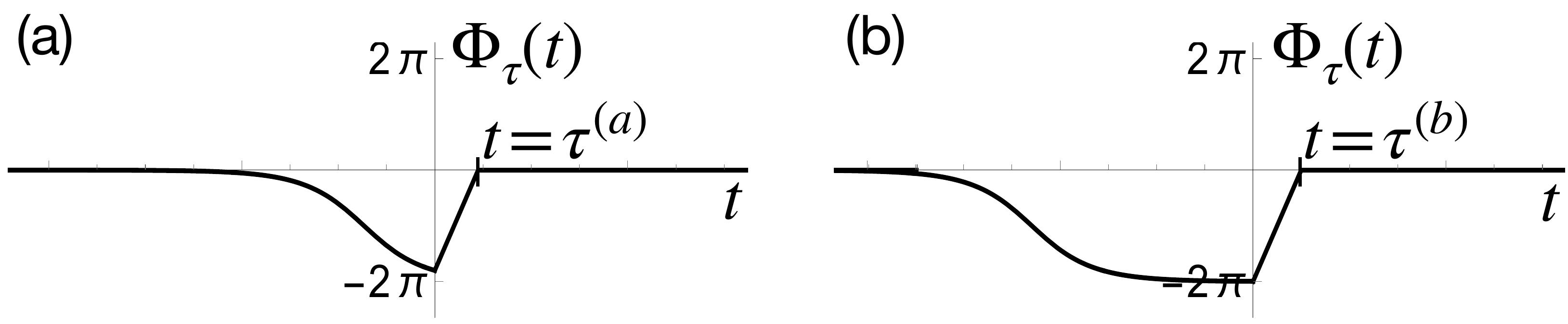}}
	\vspace{-0.2cm}
	\center{\includegraphics[width=0.95\linewidth]{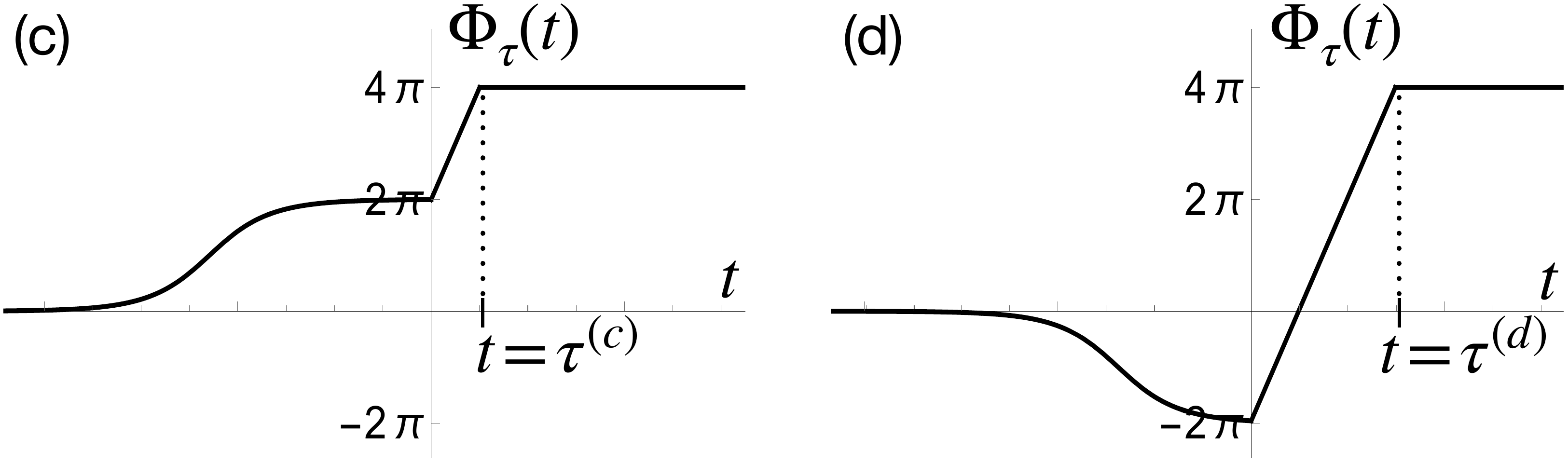}}
	\caption{   Schematic view  of the  quantum trajectory   (\ref{trajectory}). For   $t{<}0$ it has the instanton tail $\phi_\tau(t)$ (\ref{tail}),  a  linear part ${\sim }4E_{\rm c}t$   at  $0{<}t{<}\tau$, and a constant 
	  value $4\pi\mathcal{N}_\tau$ at $t{>}\tau$. 
	  Data shown for   (a)  $\tau^{(a)}{=}0.45\tau_{\rm c}$, (b)  $\tau^{(b)}{=}0.499\tau_{\rm c}$, (c) $\tau^{(c)}{=}0.501\tau_{\rm c}$, and (d)  $\tau^{(d)}{=}1.49\tau_{\rm c}$.
}
	\label{solution}
\end{figure}

{\color{black}Note the combination of  theta functions  in the r.h.s. of the equation for the quantum trajectory, $\frac{\dot\varphi_{\rm q}(t)}{8 E_{\rm c} }{-} \frac{\Gamma}{4} \sin \frac{\varphi_{\rm q} (t)}{2}{=}{-}\frac{\theta(t{-}\tau)-\theta(t)}{2}$, plays a role of an external force. It  switches on at $t{=}0$ and off at $t{=}\tau$  (assuming $\tau{>}0$). Consider first the case of    zero  force when the equation  is uniform, $ \dot\varphi_{\rm q}{=} 2 E_{\rm c}\Gamma \sin \frac{\varphi_{\rm q}}{2}$, i.e., when $t{\in}[-\infty;0]{\cup}[\tau;\infty]$. It has a set of trivial solutions, 
\be \varphi_{\rm q}{=}2\pi n \ , \label{sol-const}
\ee
 and the instanton-like ones, 
\be \varphi_{\rm q} =\pm 4{\rm arctan}(  A e^{\pi \frac{\Gamma t}{ \tau_{\rm c} } }  )+ 4\pi n \ ,\label{sol-inst}
\ee
 with $n{\in}\mathbb{Z}$. Here, the constant   $A$ determines the instanton center and the  time scale $\tau_{\rm c}$  is inverse proportional to the charging energy, $\tau_{\rm c}{=}\frac{\pi\hbar}{ E_{\rm c} }$. The instanton has slow dynamics on the long  $RC$-like time scale,  $ {\sim}\frac{\tau_{\rm c}}{\pi \Gamma  }$,  due to $\Gamma{\ll}1$.   

Consider the second case when the force is switched on ($t{\in}[0;\tau]$) and the equation becomes $\frac{\dot\varphi_{\rm q}(t)}{8 E_{\rm c} }{-} \frac{\Gamma}{4} \sin \frac{\varphi_{\rm q} (t)}{2}{=}\frac{1}{2}$. We   neglect   the sine term due to small $\Gamma{\ll}1$ prefactor and obtain  a rapidly growing linear solution 
\be \varphi_{\rm q} = 4E_{\rm c} t+B \label{sol-linear}
\ee
  up to small    oscillations with an amplitude ${\sim}\Gamma$. 

We have to match the linear solution (\ref{sol-linear}) in the region $t{\in}[0;\tau]$ with the solutions in the remaining two regions, $[-\infty;0]$  and $[\tau;\infty]$.   The analysis shows that,  in order to satisfy  the above selection rules, the solution in the region $t{<}0$ should be of the instanton form (\ref{sol-inst}) and in the other region, $t{>}\tau$, the constant one, Eq.~(\ref{sol-const}), with an even $n$.
The matching conditions (continuity of the solution)   uniquely determine free parameters $A_\tau$ and $B_\tau$, as well as the sign of the instanton function, and the even integer $ n_\tau{=}2{N}_\tau$, where we added the subscript $\tau$ to emphasize the $\tau$-dependence. 
 In particular, we have  
\be
{N}_\tau=\lfloor
   \tau/\tau_{\rm c} +1/2\rfloor
   \ee
    where the floor function $\lfloor x \rfloor $ returns the greatest integer less than or equal to   $x$. Note that the integer valued function $ {N}_\tau$ is odd, $ {N}_\tau{=}{-} {N}_{-\tau}$. 
    
    After some algebra, we find the final  result for} the quantum  trajectory at $\tau>0$:
 \begin{equation}
 \!\!\Phi_\tau(t) =   
 \! \begin{cases}
     \phi_\tau(t)\ , \ t<0; \\ 
4\pi    
 {N}_\tau{+} 4 E_{\rm c} (t{-}\tau)   ,  0{<}t{<}\tau; \\ 
4\pi   
  {N}_\tau \ ,\ t>\tau. 
 \end{cases} \!\!\!\!\!\!\!\!\!
 \label{trajectory}
\end{equation}
{\color{black}The view of the solution is shown in Fig.~\ref{solution} for different $\tau/\tau_{\rm c}$ ratios.}
 The instanton tail in (\ref{trajectory}) reads
\be
\!\!\!\!\phi_\tau(t )=4(-1)^{m_\tau}{\rm arctan}\left[  e^{\pi \frac{\Gamma t}{ \tau_{\rm c} } } \tan\!\frac{\arccos(\cos(2\pi \frac{\tau}{\tau_{\rm c}} ))}{2}\right ]\!\!\! \label{tail}
\ee
 where the discrete valued function $m_\tau$, which determines the overall  sign of (\ref{tail}), is given by $ m_\tau {=}1 {+} \big( \lfloor 
\frac{   \tau}{\tau_{\rm c}}  {+} \frac{1}{2}\rfloor {+}  \lfloor\frac{   \tau }{\tau_{\rm c}} \rfloor \big) 
$.

 Finally, one finds for the boson 
 correlator for arbitrary $\tau$:
\be
\mathcal{D}(\tau)= 2i e^{i2\pi Q_{\rm g} {N}_\tau 
}
|\!\cos(  E_{\rm c}\tau)|^{\frac{\xi   |V|}{ 2  E_{\rm c} }}   \sin\left( \frac{\pi\tau}{\tau_{\rm c}}\right) e^{-\kappa_\tau}.
 \label{exp1}
\ee
This is an oscillating   function  of  $\tau$ multiplied by a decaying envelope determined by $\kappa_\tau{=}\frac{\Gamma}{4} \xi |V|\big(|\tau|{-} \frac{1}{ 2E_{\rm c}  }\sin (2 E_{\rm c} |\tau|)\big)$. 

 Neglecting   small decay $\kappa_\tau$ in  (\ref{exp1}),   the following spectrum of excitations  is obtained,
\be
\mathcal{D}_\omega=\sum\limits_n   \mathcal{W}_\omega \delta(\omega-\omega_n) \ . 
\ee
It is a ladder of levels, $\omega_n{=}2E_{\rm c}(n{-}Q_{\rm g}{-}\frac{1}{2})$, corresponding to excitations between the states with energies $E_{n}$ and $E_{n-1}$ where $E_n=E_{\rm c}(n-Q_{\rm g})^2$ is the energy of a state with $n$ excess electrons in the island. (We note that the singularities will be slightly smoothened   by the frequency ${\sim}\Gamma E_{\rm c}$ when the $\kappa_\tau$ is taken into account.) The envelope    spectral  function $\mathcal{W}_\omega$ is
\be
\mathcal{W}_\omega=-8E_{\rm c}\int\limits_0^{\frac{\tau_{\rm c}}{2}} \!|\!\cos\!  E_{\rm c}\tau|^{\!\frac{\xi   |\!V\!|}{ 2  E_{\rm c} }}   
\sin( E_{\rm c}\tau) \sin(\omega \tau)d\tau \ . \label{W}
\ee
 It is an odd function of $\omega$, $\mathcal{W}_\omega{=}-\mathcal{W}_{-\omega}$. Its negative  (positive) 
 values correspond to rates of an absorption (emission) of an electron by a lead at the energy $\hbar \omega$.

In the high voltage regime we estimate that the weights $\mathcal{W}_{\omega_n}$ are significant up to $n{\sim} \sqrt{\frac{V}{E_{\rm c}}}$. This estimate for $n$ determines the number of the charge states that participate in the transport.

\subsection{Asymptotic expressions}
Let us come back to the expressions for the currents  (\ref{IM}) and   (\ref{ID}) and   analyze some  important cases.
We focus on the asymptotic behavior at  voltages $V{\gg}E_{\rm c} {\rm max} \{h, h^{-1}\}$ and 
integrate over  $\epsilon$ and $\omega$ analytically. In this regime the decay $\kappa_\tau$ in (\ref{exp1}) is negligible.  The following  identities  for the Fourier transformations of  distribution functions, $f_{\rm M(D)}(\tau) = \!\int \!  \frac{d\epsilon}{2\pi}e^{-i\epsilon \tau} f_{\rm M(D),\epsilon} $, are used:
$
f_{\rm M}(\tau)  {=}{-} i \frac{\gamma_{\rm L}^2\cos(\frac{1}{2}V \tau {+}\mu \tau){+} \gamma_{\rm R}^2\cos(\frac{1}{2}V \tau {-}\mu \tau)}{\pi(\gamma_{\rm L}^2+\gamma_{\rm R}^2)\tau}$
and 
$
f_{\rm D}(\tau) {=} {-}i e^{i\mu \tau}\frac{\gamma_{\rm L}^2e^{-i V \tau/2 } + \gamma_{\rm R}^2 e^{i V \tau/2 }}{\pi(\gamma_{\rm L}^2+\gamma_{\rm R}^2)\tau}. 
$
 For the difference in occupation numbers in the leads, $\Delta n_\omega{=}
 n_{\rm {L}, \omega } {-} n_{\rm {R}, \omega} $, we have  
 $
 \Delta n(\tau){=}\int\frac{d\omega}{2\pi}  e^{- i\omega \tau} \Delta n_\omega
 {=}  
 \frac{\sin\frac{V \tau}{2}}{\pi \tau}. \label{Delta}
 $
  Then    the   currents read:
  \be
I_{\rm M(D)}=gV - \pi g 
\int
e^{-i\mu\tau}\mathcal{D}(\tau) f_{\rm  M(D)}(\tau)\Delta n(-\tau)d\tau . \label{IMDtau}
\ee    The integrals in  (\ref{IMDtau})  can be split into a sum of  integrals over  intervals $\tau\in[\tau_{\rm c}(m-\frac{1}{2}); \tau_{\rm c}(m+\frac{1}{2})]$,  $m\!\in\!\mathbb{Z}$. Their integrands  have each a narrow peak in every interval.
  The peak at $m{=}0$  is given by a smoothened singularity in $\Delta n(\tau)$. In this case,  $\mathcal{D}(\tau)\!\approx\! 1$ at the relevant time scale of $\tau{\sim} \frac{1}{V}$. The integral for $m{=}0$    yields the offset current, $I_{\rm offs}$. Note that it  does not depend on $Q_{\rm g}$ and is  proportional to $ E_{\rm c} $. Integrals over the other $m{\neq} 0$ peaks    are responsible for the part of the current,  $I_{\rm osc}$, showing the gate charge oscillations.  Their amplitude  is  much smaller than that of the offset current. This means that at high voltages the strong Coulomb blockade is weakened and the charge on the island is not  well defined.  For the calculation of $I_{\rm osc}$  the boson correlator  $\mathcal{D}$  becomes important. In a $\tilde\tau $-vicinity of  $m$-th  peak, where  $\tau{=}\tau_{\rm c} m {+}\tilde\tau$, it reads    $\mathcal{D}(\tau) = 2i(-1)^{m} \sin ( E_{\rm c} 
  \tilde \tau ) e^ {i 2\pi Q_{\rm g} m}\exp\big({-}\frac{ \xi}{4} |V| E_{\rm c}  \tilde\tau^2\big)$. 
Therefore,  for both systems the current can be written  as
  \be
I_{\rm M(D)}= g V -I_{\rm M(D),offs}-I_{\rm M(D),osc}\ . \label{I}
\ee

\subsection{Offset current}
The difference between Majorana and Dirac  fermions is most prominent for asymmetric systems. We obtain the following asymptotic results ($V{\gg}E_{\rm c} {\rm max} \{\frac{\gamma_{\rm R}^2}{\gamma_{\rm L}^2} , \frac{\gamma_{\rm L}^2}{\gamma_{\rm R}^2}\}$) for the offset current. 
 In Majorana case we get 
\be
I_{\rm M, offs}= 
 g \left(1+  
 \frac{|\gamma_{\rm R}^2 -\gamma_{\rm L}^2|}{2(\gamma_{\rm R}^2+\gamma_{\rm L}^2)}\right)  E_{\rm c} \ . \label{IoffsetSymm}
\ee
In comparison, in Dirac  case we find 
\be
I_{\rm D, offs}= g  E_{\rm c}   
\ee
for an arbitrary value of $ \gamma_{\rm R}^2/\gamma_{\rm L}^2$.
Therefore, the deficit current in  the Majorana case (cf. Eq.~(\ref{IoffsetSymm})) is up to 3/2 times larger that that in the  Dirac case.

\subsection{Gate charge oscillations}
  We obtain the following asymptotic result at large voltages for Dirac case: 
\begin{multline}
\!\!I_{ \rm D,osc}{=}g  {\rm sign}(V)\frac{   \sinh \frac{2}{\xi_{\rm D} }}{  \sqrt{\pi \xi_{\rm D} }} \sqrt{\!\frac{ E_{\rm c} ^3}{ |V|} }  e^{   -\frac{  |V|}{   \xi_{\rm D}  E_{\rm c} } } \times \\
\times
 \frac{     F(\frac{V}{ 2E_{\rm c}}, Q_{\rm g}) {+} h   F(\frac{-V}{ 2E_{\rm c}}, Q_{\rm g}) }{ 1+h} .
\label{IoscDirac}
\end{multline}
There is an exponential decay of oscillations'  amplitude as a function of  $V$. 
The    gate charge oscillations pattern is given by the function $F_{x,y}{=}(2{\rm mod}_1(-x{+} y {+}\frac{1}{2}) {-}1)^2{-}\frac{1}{3}$. (The function ${\rm mod}_1(z)$ returns a fractional part of $z$.) It is found after  the integration over $\tilde \tau$  in Gaussian  approximation and further summation over $m{\neq}0$~\footnote{The sums   $\sum_{n}\frac{\cos A n}{n^2}$ or $\sum_{n}\frac{\sin A n}{n^2}$ are reduced to the polylogarithm function ${\rm Li}_2(z){=}\sum_{k=1}^\infty  \frac{z^k}{k^2}$ which has    the property ${\rm Re}[{\rm Li}_2(e^{2\pi i z}) ]{=}\pi^2 ({\rm mod}_1(z) {-}\frac{1}{2})^2{-}\frac{\pi^2}{12}$ for real $z$.}. 

The function $F$ has   discontinuous derivatives. Namely, $V$ and $Q_{\rm g}$, which satisfy the condition $F(\frac{\pm V}{ 2E_{\rm c}}, Q_{\rm g}){=}1$, determine border  lines of the so called Coulomb diamond in the differential conductance map.
As seen from (\ref{IoscDirac}), there are two sets of border lines, $Q_{\rm g}^{(1,2)}{=}{\pm} \frac{V}{2E_{\rm c}}{+}\frac{1}{2}{+}n$ ($n{\in}\mathbb{Z}$).  In asymmetric limit with $h{\ll}1$ the lines  $Q_{\rm g}^{(2)}$ are suppressed.

In Majorana case the result in general case is more cumbersome:
\begin{multline}
\!\!I_{ \rm M,osc}{=}\frac{g {\rm sign}(V)  }{2\sqrt{\pi \xi_{\rm M} } } \sqrt{\!\frac{ E_{\rm c} ^3}{ |V|} }
\!\left\{  \! \frac{(h{-}1) \sinh\frac{4\frac{\mu}{V}}{\xi_{\rm M}}}{1{+}h}  e^{\frac{-\mu^2}{E_{\rm c}|V|\xi_{\rm M}}}  F\Big(\frac{\mu}{ E_{\rm c}},\! Q_{\rm g}\!\Big){+}\!\!\!\! \right.
\\ 
 \!  \sum\limits_{j,s=\pm1}   \!\!\!  \left. h^{\frac{1-s}{2}} \frac{  \!\sinh \frac{2{+}(\!j{+}1\!)s\frac{2\mu}{V}}{\xi_{\rm M}}}{1{+}h}   e^{\!-|V|\frac{1{+}2(\!j{+}1\!)  (\!s{+}\frac{\mu}{V}\!)\frac{\mu}{V}}{E_{\rm c}\xi_{\rm M}}} F
\Big(\frac{jsV\!{+}(\!j{+}1\!)\mu}{ 2E_{\rm c}},\! Q_{\rm g}\!\Big)\right\}\!.\!\!\!\!
\label{IoscMajorana}
\end{multline}
The terms with $j{=}{-}1$ and $s{=}{\pm}1$ in the sum (\ref{IoscMajorana})   provide the same      border lines, $Q_{\rm g}^{(1,2)}{=} \frac{\pm V}{2E_{\rm c}}{+}\frac{1}{2}{+}n$, as in the Dirac case. Other   terms define   three additional sets of lines: 
$Q_{\rm g}^{(3,4)}{=} \frac{\pm V +2 \mu}{2E_{\rm c}}{+}\frac{1}{2}{+}n$ and $Q_{\rm g}^{(5)}{=} \frac{ \mu}{E_{\rm c}}{+}\frac{1}{2}{+}n$. Note that $Q_{\rm g}^{(3,4,5)}$ depend on    $\mu{=}\frac{(1-h)V}{2(1+h)}$ and, hence,  make the patterns more complicated.
We also find that in two particular cases, fully symmetric ($h{=}1$) and    absolutely asymmetric systems ($h{=}0$ or $h{=}\infty$), the patterns for 
$I_{\rm D}$ and $I_{\rm M} $  coincide. 

\begin{figure}[h!]
	\center{\includegraphics[width=\linewidth]{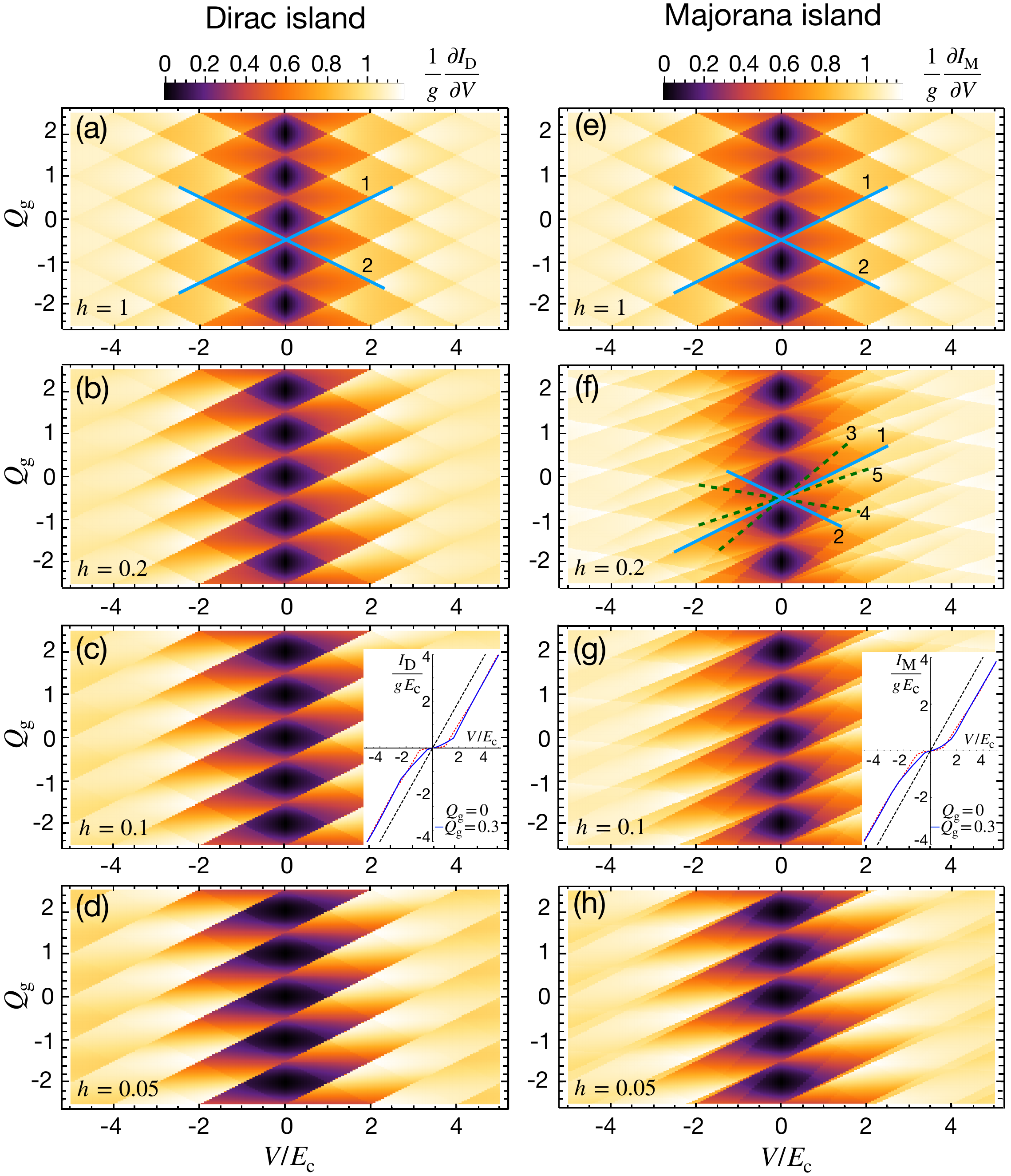}}
	\caption{ Normalized differential conductances for the Dirac and Majorana SETs, $\frac{1}{g}\frac{\partial I_{\rm D,M}}{\partial V}$,  plotted for different asymmetry parameters $h=\frac{\gamma_{\rm R}}{\gamma_{\rm L}}$. Left column   [panels (a)-(d)]: $\frac{1}{g}\frac{\partial I_{\rm D}}{\partial V}$ as  functions of $V$ and $Q_{\rm g}$ in the       Dirac   case.  Right column  [panels (e)-(h)]: $\frac{1}{g}\frac{\partial I_{\rm M}}{\partial V}$ for the Majorana island. 
	In symmetric limit $h=1$  [panels (a), (e)], the patterns are identical with the  border lines are $Q^{(1,2)}_{\rm g}{=}{\pm} V/E_{\rm c}{+}\frac{1}{2}{+}n$ ($n{\in}\mathbb{Z}$). In Dirac case with $h{\neq}1$   [panels (b)-(d)],   the lines    $Q^{(2)}_{\rm g}$  have smaller magnitude. In Majorana case  [panels (f)-(h)] the asymmetry causes  three additional  border lines $Q^{(3,4,5)}_{\rm g}$. Insets [panels (c), (g)]: asymmetric current-voltage relations   for $Q_{\rm g}=0.3$ (solid blue curves).     Red dashed curves  are the   symmetric  current-voltage relations  at $Q_{\rm g}{=}0$.
}
	\label{gDM}
\end{figure}

\subsection{Graphical presentation of the results
}

In Fig.~\ref{gDM}  we plot the normalized differential conductance  as a function of the gate charge and transport voltage for the Dirac [panels (a)-(d)] and Majorana [panels (e)-(h)] devices. Data are found after an exact integration over time in (\ref{IMDtau}). 
These plots   demonstrate  the vanishing of the border line $Q_{\rm g}^{(2)}$ at small $h$ in strongly asymmetric  Dirac devices according to asymptotic result (\ref{IoscDirac}). Also, we observe   the emerging of three additional border lines [panel (f)], $Q_{\rm g}^{(3,4,5)}$, in Majorana device at $h{\neq }1$ predicted by (\ref{IoscMajorana}).

In Fig.~\ref{gDM}  we used our formalism down to zero voltages. Quantitatively, the differential conductance at lower voltages, obtained in our formalism, is not   accurate. We are confident, however, that the pattern is qualitatively correct and reflects features of  the   strong  Coulomb  blockade  behavior.

In asymmetric junctions, the exponentaily small oscillatory contributions are not  symmetric under change of the voltage sign  $V{\to}{-}V$, i.e.,   $I_{\rm osc}(V,Q_{\rm g}){\neq}{-}I_{\rm osc}(-V,Q_{\rm g})$. The  exception is the points $Q_{\rm g}{=}\frac{n}{2}$ ($n{\in}\mathbb{Z}$) where the poles of $\mathcal{W}_\omega$ are symmetric with respect to $\omega=0$. 
This asymmetry is   more visible at low voltages, as shown in the insets in [panels (c), (g)]. It points to a possible diode-like behavior of the asymmetric  devices at low $V$.

\begin{figure}[h!]
	\center{\includegraphics[width=\linewidth]{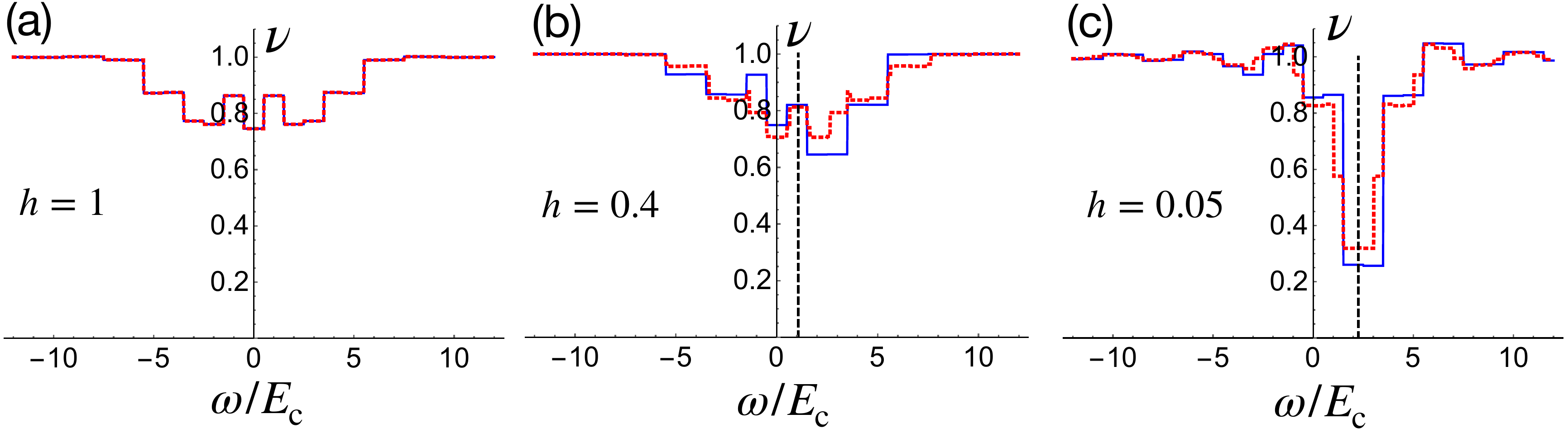}}
	\caption{   Non-equilibrium TDoS in Majorana device $\nu_{\rm M,\omega}$ (dotted red curves)  and in Dirac device $\nu_{\rm D,\omega}$ (solid blue curves). The voltage and the gate charge are chosen to be $V{=}5E_{\rm c}$ and $Q_{\rm g}{=}0$. Results are shown for (a) $h{=}1$ (symmetric device, curves match), (b)  $h{=}0.4$, and (c) $h{=}0.05$ (highly asymmetric device).  {\color{black}In Majorana case, TDoS is always symmetric around the chemical potential $\mu$ (vertical black dashed lines).}}
	 \label{dos-1}
\end{figure}

\begin{figure}[h!]
	\center{\includegraphics[width=\linewidth]{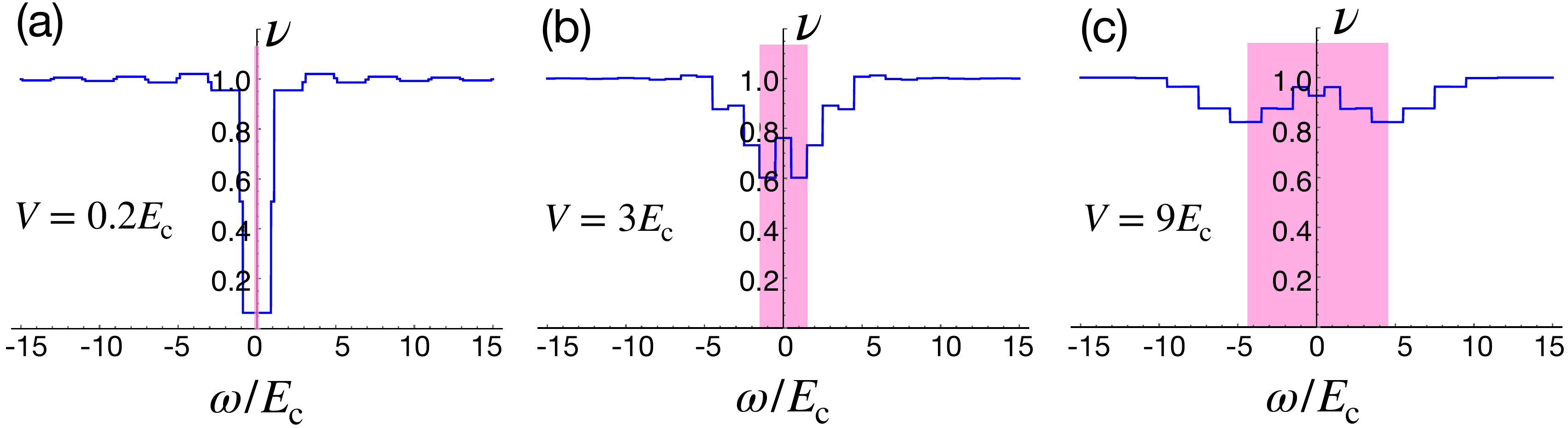}}
	\caption{  Coulomb gap structure at $\gamma_{\rm R}{=}\gamma_{\rm L}$, where $\nu{=}\nu_{\rm M}{=}\nu_{\rm D}$, for zero offset charge and different voltages: (a) $V{=}0.2E_{\rm  c}$, (b)  $V{=}3E_{\rm  c}$, and (c) $V{=}9E_{\rm  c}$. The shaded areas  are the energy windows $\frac{-eV}{2}{<}\hbar\omega{<}\frac{eV}{2}$ relevant for the charge transport.}
	 \label{dos-2}
\end{figure}

In Fig.~\ref{dos-1} we plot non-equilibrium TDoS, $\nu_{\rm M, \omega}$ and $\nu_{\rm D,\omega}$, which demonstrate  a structure of the Coulomb gap. Note that in the Majorana case we always have symmetric TDoS around the  chemical potential $\mu$ (dashed line)  for any $h$. It is dictated by the particle-hole symmetry of $f_{\rm M,\epsilon}$. In Fig.~\ref{dos-2} we demonstrate the broadening  of the Coulomb gap   when the voltage increases. Shaded regions   stand for the energy domain, the states from which contribute to the electric current.

\section{ Discussion}

The instanton-like solution presented  in Eq.~(\ref{trajectory}) provides  only      the quantum component of the phase. At the same time, the classical component is not uniquely defined and we have to integrate over  all its realizations. This is precisely the difference  from the  non-equilibrium instanton approach developed in Ref.~\cite{PhysRevB.93.155428}.  This approach is valid for   $g{\gg}1$, i.e., for the weak Coulomb blockade, and  the instanton  trajectory  fixes   both     quantum and classical phase components. 
In contrast, in our case $g{\ll}1$, the Coulomb blockade is strong   at low voltages  and is lifted at    voltages higher than the charging energy.

  Our approach based on the  AES action allows us to reproduce quantitatively some already known results   obtained within charge representation.  This is   the offset current, which  is a characteristic  feature of   charging effects at high voltages.  In the Dirac case we found $I_{\rm D, offs}{=} g  E_{\rm c}   $, which fully coincides with  the offset current in the single tunnel   junction with dimensionless conductance $g$~\cite{Averin:1986aa}. 
 Another example is the  threshold voltage  $V{=}E_{\rm c}$   for $Q_{\rm g}{=}0$ and $\gamma_{\rm R}{=}\gamma_{\rm L}$, which is two times lower than that in the ``orthodox" theory. This result can be easily obtained from Ref.~\cite{PhysRevLett.102.026805} and is due to the non-equilibrium distribution function in the dot. 
 The Coulomb blockade is lifted above this threshold. We also reproduce the threshold value $V{=}E_{\rm c}$ (see Fig.~\ref{gDM} [panel (a)]).
 
   One of the central results is   the unconventional offset current found  for the Majorana island,  $I_{\rm M, offs}{=} q_h g  E_{\rm c}   $, where the non-universal prefactor    $1{\leq}q_h{\leq}\frac{3}{2}$ {\color{black}(cf. Eq.~(\ref{qh}))} depends on the asymmetry of the SET. 
    This could serve as an evidence of the non-equilibrium chiral Majorana fermions in the island. In addition, the gate charge oscillations show distinctive features in the Majorana case, as shown in Fig.~\ref{gDM} [panel (f)]. Such measurements  could provide   an  alternative  to the  interferometry~\cite{FuKanePRL2009,PhysRevLett.102.216404,STRUBI2015489,PhysRevLett.107.136403,PhysRevB.85.125440,PhysRevB.93.155411,PhysRevB.95.195425,PhysRevB.98.245405,PhysRevB.83.100512,PhysRevB.83.220510,PhysRevB.88.075304} or time resolved transport~\cite{Lian10938, PhysRevLett.122.146803, PhysRevB.102.045431, beenakker2020shot, AdagideliSciPost2019,PhysRevB.104.035434} in Majorana devices.

\section{ Summary}
In this work, we studied the non-equilibrium transport in single-electron transistors where the strong Coulomb blockade is suppressed by large  voltages.  Two different kinds of quantum dots  were considered. These are the islands with chiral Dirac or chiral Majorana circular modes. These could be the edge states of the usual or anomalous quantum Hall insulators (Dirac) or of the proximity-induced 2D topological superconductors (Majorana). The results of this work are twofold. First, we calculated the non-equilibrium tunneling density of states and current-voltage relations. We found an unusual behavior of the offset current in the Majorana case. There are also  distinctive features  in the residual gate charge oscillations  of the transport current (Coulomb diamond) in the Majorana case.  Second, on the methodological level, we developed an instanton-like approach in the Keldysh formalism   in the limit of small conductances and high voltages.

\begin{acknowledgments}
This research was financially supported by the DFG  Grants No. MI 658/12-1, MI 658/13-1, SH 81/6-1, and by RFBR Grant No. 20-52-12034.
\end{acknowledgments}

\appendix
\section{Gauge transformation}
\label{AGT}
The Bogolyubov-de Gennes (BdG) Hamiltonian of the 2D topological insulator electrons {\color{black}($\bar c_{\uparrow,\downarrow}$, $c_{\uparrow,\downarrow}$)}  in a proximity with $s$-wave superconducting island with the pairing potential $\Delta(t)$  reads
\be \!\!
H(t){=}
\frac{1}{2}\!\sum\limits_{s,s'}\!\int\!\! d^2\mathbf{r}\begin{bmatrix}
\bar c & c\end{bmatrix}_{s}\!
\begin{bmatrix}
H_{\rm TI} {-} \mathcal{V}(t)  & i s^y \Delta(t)  
\\ \\
-is^y\Delta^*(t) & - H_{\rm TI}^T{+} \mathcal{V}(t)
\end{bmatrix}_{s,s'}\!
\begin{bmatrix}
  c \\ \\ \bar c  \end{bmatrix}_{s'}  \!\! \label{H-prox}
\ee
The Hamiltonian $H_{\rm TI}$ describes  the topological part of the island, $s$ is the spin index and $s^y$ is the Pauli matrix in the spin space.
{\color{black}The pairing potential, $\Delta(t){=}\Delta_0e^{-2i\mu t -  i \varphi(t)}$,   appears after a Hubbard-Stratonovich decoupling of an interaction term    in the superconducting island. The superconducting  phase   involves the zero mode $\mu $, which   is   the yet unknown chemical potential in the superconductor. The non-zero modes are captured by $\varphi(t)$. The potential $\mathcal{V}(t)$ appears after another Hubbard-Stratonovich transformation that decouples the charging energy in the Hamiltonian. It reads as $\mathcal{V}(t){=}\mathcal{V}_0{+}\delta \mathcal{V}(t)$ where $\mathcal{V}_0$ is its zero mode    and $\delta \mathcal{V}(t)$ involves all non-zero ones. The gauge transformation, which allows us to eliminate the phase dependence from the order parameter $\Delta(t)$ and make it real, $\Delta(t){\to}\Delta_0$, reads  as 
\be
c_s(\mathbf{r},t) {\to} e^{-i\mu t -i\frac{1}{2} \varphi(t)} c_s(\mathbf{r},t) , \ \bar c_s(\mathbf{r},t) {\to} e^{i\mu t +i\frac{1}{2} \varphi(t)} \bar c_s(\mathbf{r},t) .
\ee
 As a result, the diagonal part of the BdG Hamiltonian changes accordingly, 
\be H_{\rm TI} {-} \mathcal{V}(t) \to  H_{\rm TI} {-}  (\mathcal{V}_0-\mu)-\Big (\delta \mathcal{V}(t)- \frac{1}{2}\dot\varphi(t)\Big) .  \label{HTI}
\ee
The Anderson-Higgs mechanism   in the superconductor suppresses   the non-stationary term $\Big (\delta \mathcal{V}(t)- \frac{1}{2}\dot\varphi(t)\Big)$  in (\ref{HTI})~\cite{altland2010condensed}.  Thus, we obtain the Josephson relation between the phase and potential, $\delta \mathcal{V}(t){=} \frac{1}{2}\dot\varphi(t)$. The zero modes, $\mathcal{V}_0$ and $\mu$, are determined by   global conditions, e.g., the capacitive relation between the charge and the potential, or the conservation of current, the latter being the main subject of our calculation.  We arrive at the stationary BdG Hamiltonian (\ref{H-prox}) with $H_{\rm TI}{-}( \mathcal{V}_0{-}\mu)$ at the diagonal, which we assume to be in the superconducting topological phase with the gap $\Delta_0$. The low energy excitations of (\ref{H-prox}) are assumed to be  the chiral Majorana edge modes, $\chi$, described by the effective action (\ref{SM}).  Assuming that transport voltages are smaller than $\Delta_0$ and, possibly, the topological gap, the linear dispersion of the Majorana eigenstates is not affected by the presence of $( \mathcal{V}_0{-}\mu)$ in the BdG Hamiltonian. The gauged away phase, $\frac{1}{2}\varphi(t)+\mu t$, appears in the tunneling action~(\ref{St}).

 \section{Calculation of the Green function of the   chiral fermions in the island}
\label{GLOC}
In this Appendix we show that the local Green functions of chiral fermion at the contact points, $\mathbf{G}_{{\rm 0},\epsilon}(x_{\rm L},x_{\rm L})$ and $\mathbf{G}_{{\rm 0},\epsilon}(x_{\rm R},x_{\rm R})$, are equal and are denoted by $\mathbf{G}_{{\rm 0},\epsilon}$  in (\ref{G-loc-2}).
To derive $\mathbf{G}_{{\rm 0},\epsilon}$ explicitly,   
consider the Dyson equation for the coordinate dependent Green function of the circular chiral fermions $\mathbf{G}_{{\rm 0}}(x,t;x',t')$:
\be 
\mathbf{G}^{-1}_0\mathbf{G}_{{\rm 0}}(x,t;x',t'){=}\sigma_0 \delta(x{-}x')\delta(t{-}t'). \label{Dyson}
\ee  The stationary   integral-differential operator $\mathbf{G}^{-1}_0$, which account for the presence of the contacts,  is given by  (\ref{G0-1}).  After  the Fourier transformation, Eq.~(\ref{Dyson}) yields
\begin{equation}
\Big((\epsilon{+}iv\partial_x)\sigma_z {-} \!\!\sum\limits_{l={\rm L,R}}\delta(x{-}x_{l}) \Sigma_{l,\epsilon} \Big)\mathbf{G}_\epsilon(x,x')  
=\sigma_0 \delta(x{-}x').
\end{equation}
We introduced here the self-energies related to the  left and right tunnel contacts, $\Sigma_{l,\epsilon} {=}  \gamma_{l}^2 \sigma_z  \left[    {\mathcal G}_{{\rm L},\mu+\epsilon}     { -}     {\mathcal G}_{l, \mu- \epsilon}^T       \right] \sigma_z$, where $l{=}{\rm  L,R}$.  Substituting here the  lead's Green functions $\mathcal {G}_l$ from (\ref{G-lead}), the self-energies read
\be
\Sigma_{l,\epsilon} = -  i \frac{  \gamma_{l}^2}{2\pi v} \left((\sigma_0+\sigma_x) \frac{f_{l, \mu+\epsilon}-f_{l, \mu-\epsilon}}{2} - i\sigma_y \right)  \ . \label{Sigma-R}
\ee
The next step is the  Fourier transformation in a basis of eigenstates of an isolated Majorana edge mode of the length $L$:
\be
\mathbf{G}_{\epsilon}(k_n,k_{n'}) =  L^{-2}\int\limits_0^L dx\int\limits_0^L dx' \mathbf{G}_\epsilon(x,x') e^{-i k_n x+ik_{n'}x'}  \ . 
\ee
 The eigenstates read $e^{i k_n x}$ ($n{\in} \mathbb{Z}$) where the  wave vectors and energies are given by $k_n{=}\epsilon_n/v$ and $\epsilon_n{=}E_{\rm Th}(n{+}n_{\rm v}/2)$, respectively. Here,   $n_{\rm v}$ is an integer number, which is determined by the presence of  Berry phase and the number of vortices  in the superconductor. 
 Performing the direct and then, the inverse  Fourier transformations, we obtain two equations for $\mathbf{G}_\epsilon(x_{\rm L},k_{n'})  $ and $ \mathbf{G}_\epsilon(x_{\rm R},k_{n'})$:
 \begin{multline}
\begin{bmatrix}
\sigma_z {-}     g_\epsilon(0)  \Sigma_{\rm L, \epsilon} &  {-}   g_\epsilon(x_{\rm L}{-} x_{\rm R})    \Sigma_{\rm R, \epsilon}  \\ \\
 {-}    g_\epsilon(x_{\rm R} {-} x_{\rm L})     \Sigma_{\rm L, \epsilon}  &  \sigma_0 {-}     g_\epsilon(0)  \Sigma_{\rm R, \epsilon} 
\end{bmatrix}
\begin{bmatrix}
\mathbf{G}_\epsilon(x_{\rm L},k_{n'})  \\ \\
 \mathbf{G}_\epsilon(x_{\rm R},k_{n'})
\end{bmatrix}
=\\
=\frac{1}{L(\epsilon- \epsilon_{n'})}
\begin{bmatrix}
e^{i  k_{n'} x_{\rm L}}\sigma_0  \\ \\
e^{i k_{n'}x_{\rm R}}\ \sigma_0
\end{bmatrix}
\ . \label{G-loc-0}
\end{multline}
The function $g_\epsilon(x){=}\frac{1}{L}\sum_n\frac{e^{i k_n x}}{\epsilon-\epsilon_n}$ has been introduced.
 We need to obtain the expressions for  $\mathbf{G}_\epsilon(x_{\rm L},x_{\rm L}){=} \sum_{n}  \mathbf{G}_\epsilon(x_{\rm L},k_{n}) e^{-ik_{n}x_{\rm L}}$ and $
 \mathbf{G}_\epsilon(x_{\rm R},x_{\rm R}){=}\sum_{n}  \mathbf{G}_\epsilon(x_{\rm R},k_{n}) e^{-ik_{n}x_{\rm R}} $. They follow from (\ref{G-loc-0}):
  \begin{multline}
\begin{bmatrix}
\mathbf{G}_\epsilon(x_{\rm L},x_{\rm L})  \\ \\
 \mathbf{G}_\epsilon(x_{\rm R},x_{\rm R})
\end{bmatrix}
= \sum\limits_{n}
\frac{1}{L(\epsilon{-} \epsilon_{n})}
\begin{bmatrix}
e^{-i  k_{n} x_{\rm L}}& 0  \\ \\
0 & e^{-i k_{n}x_{\rm R}}
\end{bmatrix}\times \\ 
\begin{bmatrix}
\sigma_z {-}    g_\epsilon(0)  \Sigma_{\rm L, \epsilon} &  {-}  g_\epsilon(x_{\rm L} {-} x_{\rm R})    \Sigma_{\rm R, \epsilon}  \\ \\
 -    g_\epsilon(x_{\rm R} {-} x_{\rm L})    \Sigma_{\rm L, \epsilon}  &  \sigma_z {-}    g_\epsilon(0)  \Sigma_{\rm R, \epsilon} 
\end{bmatrix}^{-1}
\begin{bmatrix}
 e^{i  k_{n} x_{\rm L}}\sigma_0  \\ \\
 e^{i k_{n}x_{\rm R}}\sigma_0
\end{bmatrix}
\ . \label{G-loc-1}
  \end{multline}
 The obtained Green functions are given by a  sequence of peaks near $\epsilon{=}\epsilon_n$. In the limit  when the peak width, ${\sim} \frac{\sqrt{\gamma_{\rm L}^2 + \gamma_{\rm R}^2}}{L}$, is much smaller than the level spacing, $E_{\rm Th}$, these can be replaced by delta functions. This condition of small peak broadening is equivalent to the tunnel approximation, $\gamma_{\rm L,R}{\ll}v$. To obtain the result in this limit, one has to drop all terms in the sum in $g_\epsilon(x)$ except those $n$ which correspond to $\epsilon_n$ being in the vicinity of $\epsilon$, i.e.,  the following replacement, $g_\epsilon(x)\to \frac{e^{i k_n x}}{L(\epsilon-  \epsilon_{n})}$. Reducing the result of the matrix inversion in (\ref{G-loc-1}) to a Lorentzian form and  approximating    the Lorentzians by  the delta functions, we obtain the result (\ref{G-loc-2}):
\begin{multline}
\mathbf{G}_\epsilon(x_{\rm L},x_{\rm L}){=} \mathbf{G}_\epsilon(x_{\rm R},x_{\rm R})=\\
\frac{ - i}{2\pi v}  \sum\limits_n\frac{\gamma_{\rm L}^2 + \gamma_{\rm R}^2}{L^2(\epsilon-\epsilon_n)^2+\frac{(\gamma_{\rm L}^2 + \gamma_{\rm R}^2)^2}{4\pi^2 v^2}}
\Big ((\sigma_0 - \sigma_x ) f_{\rm M, \epsilon} + i \sigma_y\Big) \to \\
-i \frac{E_{\rm Th}}{2v}  \sum\limits_n\delta(\epsilon-\epsilon_n) \Big ((\sigma_0 - \sigma_x ) f_{\rm M, \epsilon} + i \sigma_y\Big) 
\end{multline}
  where the   non-Fermi distribution function of the Majorana fermions reads
\be
f_{{\rm M},\epsilon}=\frac{ \gamma_{\rm L}^2(f_{{\rm L}, \mu+\epsilon}-f_{{\rm L}, \mu-\epsilon})+\gamma_{\rm R}^2(f_{{\rm R}, \mu+\epsilon}-f_{{\rm R}, \mu-\epsilon})}{2  \left(\gamma_{\rm L}^2+\gamma_{\rm R}^2\right)} \ . 
\ee
 }

 \section{Derivation  of the  AES action}
\label{ATR}
 The derivation of the AES action~(\ref{Kd}) from (\ref{Sdiss})  involves   the  following  transformation under the trace 
\begin{multline}
{\rm tr}\Big[\mathbf{G}_{\rm 0 } \Sigma[\varphi , \eta{=}0] \Big]{=}\\
\!\!\int\!\!\! dtdt'{\rm tr}_\sigma\Big[\gamma_l^2\mathbf{G}_{{\rm 0},{t'\!-t}}(U^+_t \sigma^z   {\mathcal G}_{l,t{-}t'} \sigma^z   U_{t'}{-}U_t \sigma^z   {\mathcal G}^T_{l,t'{-}t} \sigma^z   U^+_{t'})\Big]{=} \\
\!\!2(\gamma_{\rm L}^2{+}\gamma_{\rm R}^2)\!\!\int\!\!\! dtdt'{\rm tr}_\sigma[ \mathbf{G}_{{\rm 0},{t'\!-t}} U^+_t \! \sigma^z \!  \frac{\gamma_{\rm L}^2{\mathcal G}_{{\rm L},t{-}t'}{+}\gamma_{\rm R}^2{\mathcal G}_{{\rm R},t{-}t'}\!}{\gamma_{\rm L}^2+\gamma_{\rm R}^2} \sigma^z   U_{t'}\!]. \!\!\!\!\! \label{TR-0}
\end{multline}
We substitute here the Majorana Green function $\mathbf{G}_{{\rm 0}}$ from (\ref{G-loc-2}) and lead's Green function $\mathcal {G}_l$ from (\ref{G-lead}), as well as the matrix $U_t{=}U[\varphi(t),\eta_l {=}0]$ from (\ref{U}) (that depends on the field $\varphi(t)$).
The symmetry of the Majorana Green function, $[\mathbf{G}_{{\rm 0},{-t}}]^T{=}{-}\mathbf{G}_{{\rm 0},{t}}$, has been exploited here. {\color{black}To obtain    expression~(\ref{Sdiss})   we subtract    the divergent stationary part ${\rm tr}\Big[\mathbf{G}_{\rm 0 } \Sigma[\varphi{=}\varphi_0, \eta{=}0] \Big]$    from (\ref{TR-0}).}  As a result, after the Keldysh rotation, we obtain the AES action~(\ref{Kd})  with   $\alpha^{R(A)}$ and $\alpha^K$ defined by Eqs.~(\ref{alphaR}) and (\ref{alphaK}). The distributions $f_{\rm M}$ and $f_{\rm D}$, which determine the kernels  $\alpha^R$ and $\alpha^K$, originate  from the dot's and leads' Green functions, $\mathbf{G}_{\rm 0}$  and   $\mathcal{G}_{\rm  L,R}$.


\begin{thebibliography}{48}%
\makeatletter
\providecommand \@ifxundefined [1]{%
 \@ifx{#1\undefined}
}%
\providecommand \@ifnum [1]{%
 \ifnum #1\expandafter \@firstoftwo
 \else \expandafter \@secondoftwo
 \fi
}%
\providecommand \@ifx [1]{%
 \ifx #1\expandafter \@firstoftwo
 \else \expandafter \@secondoftwo
 \fi
}%
\providecommand \natexlab [1]{#1}%
\providecommand \enquote  [1]{``#1''}%
\providecommand \bibnamefont  [1]{#1}%
\providecommand \bibfnamefont [1]{#1}%
\providecommand \citenamefont [1]{#1}%
\providecommand \href@noop [0]{\@secondoftwo}%
\providecommand \href [0]{\begingroup \@sanitize@url \@href}%
\providecommand \@href[1]{\@@startlink{#1}\@@href}%
\providecommand \@@href[1]{\endgroup#1\@@endlink}%
\providecommand \@sanitize@url [0]{\catcode `\\12\catcode `\$12\catcode
  `\&12\catcode `\#12\catcode `\^12\catcode `\_12\catcode `\%12\relax}%
\providecommand \@@startlink[1]{}%
\providecommand \@@endlink[0]{}%
\providecommand \url  [0]{\begingroup\@sanitize@url \@url }%
\providecommand \@url [1]{\endgroup\@href {#1}{\urlprefix }}%
\providecommand \urlprefix  [0]{URL }%
\providecommand \Eprint [0]{\href }%
\providecommand \doibase [0]{http://dx.doi.org/}%
\providecommand \selectlanguage [0]{\@gobble}%
\providecommand \bibinfo  [0]{\@secondoftwo}%
\providecommand \bibfield  [0]{\@secondoftwo}%
\providecommand \translation [1]{[#1]}%
\providecommand \BibitemOpen [0]{}%
\providecommand \bibitemStop [0]{}%
\providecommand \bibitemNoStop [0]{.\EOS\space}%
\providecommand \EOS [0]{\spacefactor3000\relax}%
\providecommand \BibitemShut  [1]{\csname bibitem#1\endcsname}%
\let\auto@bib@innerbib\@empty
\bibitem [{\citenamefont {Kulik}\ and\ \citenamefont
  {Shekhter}(1975)}]{kulik1975kinetic}%
  \BibitemOpen
  \bibfield  {author} {\bibinfo {author} {\bibfnamefont {I.}~\bibnamefont
  {Kulik}}\ and\ \bibinfo {author} {\bibfnamefont {R.}~\bibnamefont
  {Shekhter}},\ }\href@noop {} {\bibfield  {journal} {\bibinfo  {journal}
  {Zhur. Eksper. Teoret. Fiziki}\ }\textbf {\bibinfo {volume} {68}},\ \bibinfo
  {pages} {623} (\bibinfo {year} {1975})}\BibitemShut {NoStop}%
\bibitem [{\citenamefont {Averin}\ and\ \citenamefont
  {Likharev}(1986)}]{Averin:1986aa}%
  \BibitemOpen
  \bibfield  {author} {\bibinfo {author} {\bibfnamefont {D.~V.}\ \bibnamefont
  {Averin}}\ and\ \bibinfo {author} {\bibfnamefont {K.~K.}\ \bibnamefont
  {Likharev}},\ }\href {\doibase 10.1007/BF00683469} {\bibfield  {journal}
  {\bibinfo  {journal} {Journal of Low Temperature Physics}\ }\textbf {\bibinfo
  {volume} {62}},\ \bibinfo {pages} {345} (\bibinfo {year} {1986})}\BibitemShut
  {NoStop}%
\bibitem [{\citenamefont {Ingold}\ and\ \citenamefont
  {Nazarov}(1992)}]{ingold1992charge}%
  \BibitemOpen
  \bibfield  {author} {\bibinfo {author} {\bibfnamefont {G.-L.}\ \bibnamefont
  {Ingold}}\ and\ \bibinfo {author} {\bibfnamefont {Y.~V.}\ \bibnamefont
  {Nazarov}},\ }in\ \href@noop {} {\emph {\bibinfo {booktitle} {Single charge
  tunneling}}}\ (\bibinfo  {publisher} {Springer},\ \bibinfo {year} {1992})\
  pp.\ \bibinfo {pages} {21--107}\BibitemShut {NoStop}%
\bibitem [{\citenamefont {Kamenev}\ and\ \citenamefont
  {Gefen}(1996)}]{PhysRevB.54.5428}%
  \BibitemOpen
  \bibfield  {author} {\bibinfo {author} {\bibfnamefont {A.}~\bibnamefont
  {Kamenev}}\ and\ \bibinfo {author} {\bibfnamefont {Y.}~\bibnamefont
  {Gefen}},\ }\href {\doibase 10.1103/PhysRevB.54.5428} {\bibfield  {journal}
  {\bibinfo  {journal} {Phys. Rev. B}\ }\textbf {\bibinfo {volume} {54}},\
  \bibinfo {pages} {5428} (\bibinfo {year} {1996})}\BibitemShut {NoStop}%
\bibitem [{\citenamefont {Kouwenhoven}\ \emph {et~al.}(1997)\citenamefont
  {Kouwenhoven}, \citenamefont {Sch{\"o}n},\ and\ \citenamefont
  {Sohn}}]{Kouwenhoven1997}%
  \BibitemOpen
  \bibfield  {author} {\bibinfo {author} {\bibfnamefont {L.~P.}\ \bibnamefont
  {Kouwenhoven}}, \bibinfo {author} {\bibfnamefont {G.}~\bibnamefont
  {Sch{\"o}n}}, \ and\ \bibinfo {author} {\bibfnamefont {L.~L.}\ \bibnamefont
  {Sohn}},\ }\enquote {\bibinfo {title} {Introduction to mesoscopic electron
  transport},}\ in\ \href {\doibase 10.1007/978-94-015-8839-3_1} {\emph
  {\bibinfo {booktitle} {Mesoscopic Electron Transport}}},\ \bibinfo {editor}
  {edited by\ \bibinfo {editor} {\bibfnamefont {L.~L.}\ \bibnamefont {Sohn}},
  \bibinfo {editor} {\bibfnamefont {L.~P.}\ \bibnamefont {Kouwenhoven}}, \ and\
  \bibinfo {editor} {\bibfnamefont {G.}~\bibnamefont {Sch{\"o}n}}}\ (\bibinfo
  {publisher} {Springer Netherlands},\ \bibinfo {address} {Dordrecht},\
  \bibinfo {year} {1997})\ pp.\ \bibinfo {pages} {1--44}\BibitemShut {NoStop}%
\bibitem [{\citenamefont {Aleiner}\ \emph {et~al.}(2002)\citenamefont
  {Aleiner}, \citenamefont {Brouwer},\ and\ \citenamefont
  {Glazman}}]{ALEINER2002309}%
  \BibitemOpen
  \bibfield  {author} {\bibinfo {author} {\bibfnamefont {I.}~\bibnamefont
  {Aleiner}}, \bibinfo {author} {\bibfnamefont {P.}~\bibnamefont {Brouwer}}, \
  and\ \bibinfo {author} {\bibfnamefont {L.}~\bibnamefont {Glazman}},\ }\href
  {\doibase https://doi.org/10.1016/S0370-1573(01)00063-1} {\bibfield
  {journal} {\bibinfo  {journal} {Physics Reports}\ }\textbf {\bibinfo {volume}
  {358}},\ \bibinfo {pages} {309} (\bibinfo {year} {2002})}\BibitemShut
  {NoStop}%
\bibitem [{\citenamefont {Nazarov}(1999)}]{PhysRevLett.82.1245}%
  \BibitemOpen
  \bibfield  {author} {\bibinfo {author} {\bibfnamefont {Y.~V.}\ \bibnamefont
  {Nazarov}},\ }\href {\doibase 10.1103/PhysRevLett.82.1245} {\bibfield
  {journal} {\bibinfo  {journal} {Phys. Rev. Lett.}\ }\textbf {\bibinfo
  {volume} {82}},\ \bibinfo {pages} {1245} (\bibinfo {year}
  {1999})}\BibitemShut {NoStop}%
\bibitem [{\citenamefont {Altland}\ \emph {et~al.}(2006)\citenamefont
  {Altland}, \citenamefont {Glazman}, \citenamefont {Kamenev},\ and\
  \citenamefont {Meyer}}]{ALTLAND20062566}%
  \BibitemOpen
  \bibfield  {author} {\bibinfo {author} {\bibfnamefont {A.}~\bibnamefont
  {Altland}}, \bibinfo {author} {\bibfnamefont {L.}~\bibnamefont {Glazman}},
  \bibinfo {author} {\bibfnamefont {A.}~\bibnamefont {Kamenev}}, \ and\
  \bibinfo {author} {\bibfnamefont {J.}~\bibnamefont {Meyer}},\ }\href
  {\doibase https://doi.org/10.1016/j.aop.2005.12.012} {\bibfield  {journal}
  {\bibinfo  {journal} {Annals of Physics}\ }\textbf {\bibinfo {volume}
  {321}},\ \bibinfo {pages} {2566} (\bibinfo {year} {2006})}\BibitemShut
  {NoStop}%
\bibitem [{\citenamefont {Sedlmayr}\ \emph {et~al.}(2006)\citenamefont
  {Sedlmayr}, \citenamefont {Yurkevich},\ and\ \citenamefont
  {Lerner}}]{sedlmayr2006tunnelling}%
  \BibitemOpen
  \bibfield  {author} {\bibinfo {author} {\bibfnamefont {N.}~\bibnamefont
  {Sedlmayr}}, \bibinfo {author} {\bibfnamefont {I.~V.}\ \bibnamefont
  {Yurkevich}}, \ and\ \bibinfo {author} {\bibfnamefont {I.~V.}\ \bibnamefont
  {Lerner}},\ }\href@noop {} {\bibfield  {journal} {\bibinfo  {journal} {EPL
  (Europhysics Letters)}\ }\textbf {\bibinfo {volume} {76}},\ \bibinfo {pages}
  {109} (\bibinfo {year} {2006})}\BibitemShut {NoStop}%
\bibitem [{\citenamefont {Altland}\ and\ \citenamefont
  {Egger}(2009)}]{PhysRevLett.102.026805}%
  \BibitemOpen
  \bibfield  {author} {\bibinfo {author} {\bibfnamefont {A.}~\bibnamefont
  {Altland}}\ and\ \bibinfo {author} {\bibfnamefont {R.}~\bibnamefont
  {Egger}},\ }\href {\doibase 10.1103/PhysRevLett.102.026805} {\bibfield
  {journal} {\bibinfo  {journal} {Phys. Rev. Lett.}\ }\textbf {\bibinfo
  {volume} {102}},\ \bibinfo {pages} {026805} (\bibinfo {year}
  {2009})}\BibitemShut {NoStop}%
\bibitem [{\citenamefont {Rodionov}\ \emph {et~al.}(2010)\citenamefont
  {Rodionov}, \citenamefont {Burmistrov},\ and\ \citenamefont
  {Chtchelkatchev}}]{PhysRevB.82.155317}%
  \BibitemOpen
  \bibfield  {author} {\bibinfo {author} {\bibfnamefont {Y.~I.}\ \bibnamefont
  {Rodionov}}, \bibinfo {author} {\bibfnamefont {I.~S.}\ \bibnamefont
  {Burmistrov}}, \ and\ \bibinfo {author} {\bibfnamefont {N.~M.}\ \bibnamefont
  {Chtchelkatchev}},\ }\href {\doibase 10.1103/PhysRevB.82.155317} {\bibfield
  {journal} {\bibinfo  {journal} {Phys. Rev. B}\ }\textbf {\bibinfo {volume}
  {82}},\ \bibinfo {pages} {155317} (\bibinfo {year} {2010})}\BibitemShut
  {NoStop}%
\bibitem [{\citenamefont {Zazunov}\ \emph {et~al.}(2011)\citenamefont
  {Zazunov}, \citenamefont {Yeyati},\ and\ \citenamefont
  {Egger}}]{PhysRevB.84.165440}%
  \BibitemOpen
  \bibfield  {author} {\bibinfo {author} {\bibfnamefont {A.}~\bibnamefont
  {Zazunov}}, \bibinfo {author} {\bibfnamefont {A.~L.}\ \bibnamefont {Yeyati}},
  \ and\ \bibinfo {author} {\bibfnamefont {R.}~\bibnamefont {Egger}},\ }\href
  {\doibase 10.1103/PhysRevB.84.165440} {\bibfield  {journal} {\bibinfo
  {journal} {Phys. Rev. B}\ }\textbf {\bibinfo {volume} {84}},\ \bibinfo
  {pages} {165440} (\bibinfo {year} {2011})}\BibitemShut {NoStop}%
\bibitem [{\citenamefont {Shnirman}\ \emph {et~al.}(2015)\citenamefont
  {Shnirman}, \citenamefont {Gefen}, \citenamefont {Saha}, \citenamefont
  {Burmistrov}, \citenamefont {Kiselev},\ and\ \citenamefont
  {Altland}}]{PhysRevLett.114.176806}%
  \BibitemOpen
  \bibfield  {author} {\bibinfo {author} {\bibfnamefont {A.}~\bibnamefont
  {Shnirman}}, \bibinfo {author} {\bibfnamefont {Y.}~\bibnamefont {Gefen}},
  \bibinfo {author} {\bibfnamefont {A.}~\bibnamefont {Saha}}, \bibinfo {author}
  {\bibfnamefont {I.~S.}\ \bibnamefont {Burmistrov}}, \bibinfo {author}
  {\bibfnamefont {M.~N.}\ \bibnamefont {Kiselev}}, \ and\ \bibinfo {author}
  {\bibfnamefont {A.}~\bibnamefont {Altland}},\ }\href {\doibase
  10.1103/PhysRevLett.114.176806} {\bibfield  {journal} {\bibinfo  {journal}
  {Phys. Rev. Lett.}\ }\textbf {\bibinfo {volume} {114}},\ \bibinfo {pages}
  {176806} (\bibinfo {year} {2015})}\BibitemShut {NoStop}%
\bibitem [{\citenamefont {Albrecht}\ \emph {et~al.}(2016)\citenamefont
  {Albrecht}, \citenamefont {Higginbotham}, \citenamefont {Madsen},
  \citenamefont {Kuemmeth}, \citenamefont {Jespersen}, \citenamefont
  {Nyg{\aa}rd}, \citenamefont {Krogstrup},\ and\ \citenamefont
  {Marcus}}]{Albrecht:2016aa}%
  \BibitemOpen
  \bibfield  {author} {\bibinfo {author} {\bibfnamefont {S.~M.}\ \bibnamefont
  {Albrecht}}, \bibinfo {author} {\bibfnamefont {A.~P.}\ \bibnamefont
  {Higginbotham}}, \bibinfo {author} {\bibfnamefont {M.}~\bibnamefont
  {Madsen}}, \bibinfo {author} {\bibfnamefont {F.}~\bibnamefont {Kuemmeth}},
  \bibinfo {author} {\bibfnamefont {T.~S.}\ \bibnamefont {Jespersen}}, \bibinfo
  {author} {\bibfnamefont {J.}~\bibnamefont {Nyg{\aa}rd}}, \bibinfo {author}
  {\bibfnamefont {P.}~\bibnamefont {Krogstrup}}, \ and\ \bibinfo {author}
  {\bibfnamefont {C.~M.}\ \bibnamefont {Marcus}},\ }\href {\doibase
  10.1038/nature17162} {\bibfield  {journal} {\bibinfo  {journal} {Nature}\
  }\textbf {\bibinfo {volume} {531}},\ \bibinfo {pages} {206} (\bibinfo {year}
  {2016})}\BibitemShut {NoStop}%
\bibitem [{\citenamefont {Titov}\ and\ \citenamefont
  {Gutman}(2016)}]{PhysRevB.93.155428}%
  \BibitemOpen
  \bibfield  {author} {\bibinfo {author} {\bibfnamefont {M.}~\bibnamefont
  {Titov}}\ and\ \bibinfo {author} {\bibfnamefont {D.~B.}\ \bibnamefont
  {Gutman}},\ }\href {\doibase 10.1103/PhysRevB.93.155428} {\bibfield
  {journal} {\bibinfo  {journal} {Phys. Rev. B}\ }\textbf {\bibinfo {volume}
  {93}},\ \bibinfo {pages} {155428} (\bibinfo {year} {2016})}\BibitemShut
  {NoStop}%
\bibitem [{\citenamefont {Ludwig}\ \emph {et~al.}(2017)\citenamefont {Ludwig},
  \citenamefont {Burmistrov}, \citenamefont {Gefen},\ and\ \citenamefont
  {Shnirman}}]{PhysRevB.95.075425}%
  \BibitemOpen
  \bibfield  {author} {\bibinfo {author} {\bibfnamefont {T.}~\bibnamefont
  {Ludwig}}, \bibinfo {author} {\bibfnamefont {I.~S.}\ \bibnamefont
  {Burmistrov}}, \bibinfo {author} {\bibfnamefont {Y.}~\bibnamefont {Gefen}}, \
  and\ \bibinfo {author} {\bibfnamefont {A.}~\bibnamefont {Shnirman}},\ }\href
  {\doibase 10.1103/PhysRevB.95.075425} {\bibfield  {journal} {\bibinfo
  {journal} {Phys. Rev. B}\ }\textbf {\bibinfo {volume} {95}},\ \bibinfo
  {pages} {075425} (\bibinfo {year} {2017})}\BibitemShut {NoStop}%
\bibitem [{\citenamefont {Pikulin}\ \emph {et~al.}(2019)\citenamefont
  {Pikulin}, \citenamefont {Flensberg}, \citenamefont {Glazman}, \citenamefont
  {Houzet},\ and\ \citenamefont {Lutchyn}}]{PhysRevLett.122.016801}%
  \BibitemOpen
  \bibfield  {author} {\bibinfo {author} {\bibfnamefont {D.}~\bibnamefont
  {Pikulin}}, \bibinfo {author} {\bibfnamefont {K.}~\bibnamefont {Flensberg}},
  \bibinfo {author} {\bibfnamefont {L.~I.}\ \bibnamefont {Glazman}}, \bibinfo
  {author} {\bibfnamefont {M.}~\bibnamefont {Houzet}}, \ and\ \bibinfo {author}
  {\bibfnamefont {R.~M.}\ \bibnamefont {Lutchyn}},\ }\href {\doibase
  10.1103/PhysRevLett.122.016801} {\bibfield  {journal} {\bibinfo  {journal}
  {Phys. Rev. Lett.}\ }\textbf {\bibinfo {volume} {122}},\ \bibinfo {pages}
  {016801} (\bibinfo {year} {2019})}\BibitemShut {NoStop}%
\bibitem [{\citenamefont {Burmistrov}\ \emph {et~al.}(2020)\citenamefont
  {Burmistrov}, \citenamefont {Gefen}, \citenamefont {Shapiro},\ and\
  \citenamefont {Shnirman}}]{PhysRevLett.124.196801}%
  \BibitemOpen
  \bibfield  {author} {\bibinfo {author} {\bibfnamefont {I.~S.}\ \bibnamefont
  {Burmistrov}}, \bibinfo {author} {\bibfnamefont {Y.}~\bibnamefont {Gefen}},
  \bibinfo {author} {\bibfnamefont {D.~S.}\ \bibnamefont {Shapiro}}, \ and\
  \bibinfo {author} {\bibfnamefont {A.}~\bibnamefont {Shnirman}},\ }\href
  {\doibase 10.1103/PhysRevLett.124.196801} {\bibfield  {journal} {\bibinfo
  {journal} {Phys. Rev. Lett.}\ }\textbf {\bibinfo {volume} {124}},\ \bibinfo
  {pages} {196801} (\bibinfo {year} {2020})}\BibitemShut {NoStop}%
\bibitem [{\citenamefont {Ludwig}\ \emph {et~al.}(2020)\citenamefont {Ludwig},
  \citenamefont {Burmistrov}, \citenamefont {Gefen},\ and\ \citenamefont
  {Shnirman}}]{PhysRevResearch.2.023221}%
  \BibitemOpen
  \bibfield  {author} {\bibinfo {author} {\bibfnamefont {T.}~\bibnamefont
  {Ludwig}}, \bibinfo {author} {\bibfnamefont {I.~S.}\ \bibnamefont
  {Burmistrov}}, \bibinfo {author} {\bibfnamefont {Y.}~\bibnamefont {Gefen}}, \
  and\ \bibinfo {author} {\bibfnamefont {A.}~\bibnamefont {Shnirman}},\ }\href
  {\doibase 10.1103/PhysRevResearch.2.023221} {\bibfield  {journal} {\bibinfo
  {journal} {Phys. Rev. Res.}\ }\textbf {\bibinfo {volume} {2}},\ \bibinfo
  {pages} {023221} (\bibinfo {year} {2020})}\BibitemShut {NoStop}%
\bibitem [{\citenamefont {Dotdaev}\ \emph {et~al.}(2021)\citenamefont
  {Dotdaev}, \citenamefont {Rodionov},\ and\ \citenamefont
  {Tikhonov}}]{DOTDAEV2021127736}%
  \BibitemOpen
  \bibfield  {author} {\bibinfo {author} {\bibfnamefont {A.}~\bibnamefont
  {Dotdaev}}, \bibinfo {author} {\bibfnamefont {Y.}~\bibnamefont {Rodionov}}, \
  and\ \bibinfo {author} {\bibfnamefont {K.}~\bibnamefont {Tikhonov}},\ }\href
  {\doibase https://doi.org/10.1016/j.physleta.2021.127736} {\bibfield
  {journal} {\bibinfo  {journal} {Physics Letters A}\ }\textbf {\bibinfo
  {volume} {419}},\ \bibinfo {pages} {127736} (\bibinfo {year}
  {2021})}\BibitemShut {NoStop}%
\bibitem [{\citenamefont {Ambegaokar}\ \emph {et~al.}(1982)\citenamefont
  {Ambegaokar}, \citenamefont {Eckern},\ and\ \citenamefont
  {Sch\"on}}]{PhysRevLett.48.1745}%
  \BibitemOpen
  \bibfield  {author} {\bibinfo {author} {\bibfnamefont {V.}~\bibnamefont
  {Ambegaokar}}, \bibinfo {author} {\bibfnamefont {U.}~\bibnamefont {Eckern}},
  \ and\ \bibinfo {author} {\bibfnamefont {G.}~\bibnamefont {Sch\"on}},\ }\href
  {\doibase 10.1103/PhysRevLett.48.1745} {\bibfield  {journal} {\bibinfo
  {journal} {Phys. Rev. Lett.}\ }\textbf {\bibinfo {volume} {48}},\ \bibinfo
  {pages} {1745} (\bibinfo {year} {1982})}\BibitemShut {NoStop}%
\bibitem [{\citenamefont {Eckern}\ \emph {et~al.}(1984)\citenamefont {Eckern},
  \citenamefont {Sch\"on},\ and\ \citenamefont
  {Ambegaokar}}]{PhysRevB.30.6419}%
  \BibitemOpen
  \bibfield  {author} {\bibinfo {author} {\bibfnamefont {U.}~\bibnamefont
  {Eckern}}, \bibinfo {author} {\bibfnamefont {G.}~\bibnamefont {Sch\"on}}, \
  and\ \bibinfo {author} {\bibfnamefont {V.}~\bibnamefont {Ambegaokar}},\
  }\href {\doibase 10.1103/PhysRevB.30.6419} {\bibfield  {journal} {\bibinfo
  {journal} {Phys. Rev. B}\ }\textbf {\bibinfo {volume} {30}},\ \bibinfo
  {pages} {6419} (\bibinfo {year} {1984})}\BibitemShut {NoStop}%
\bibitem [{\citenamefont {Korshunov}(1987)}]{korshunov1987coherent}%
  \BibitemOpen
  \bibfield  {author} {\bibinfo {author} {\bibfnamefont {S.}~\bibnamefont
  {Korshunov}},\ }\href@noop {} {\bibfield  {journal} {\bibinfo  {journal}
  {JETP Lett}\ }\textbf {\bibinfo {volume} {45}} (\bibinfo {year}
  {1987})}\BibitemShut {NoStop}%
\bibitem [{\citenamefont {Keldysh}(1965)}]{keldysh1965diagram}%
  \BibitemOpen
  \bibfield  {author} {\bibinfo {author} {\bibfnamefont {L.~V.}\ \bibnamefont
  {Keldysh}},\ }\href@noop {} {\bibfield  {journal} {\bibinfo  {journal} {Sov.
  Phys. JETP}\ }\textbf {\bibinfo {volume} {20}},\ \bibinfo {pages} {1018}
  (\bibinfo {year} {1965})}\BibitemShut {NoStop}%
\bibitem [{\citenamefont {Kamenev}(2011)}]{Kamenev}%
  \BibitemOpen
  \bibfield  {author} {\bibinfo {author} {\bibfnamefont {A.}~\bibnamefont
  {Kamenev}},\ }\href@noop {} {\emph {\bibinfo {title} {Field Theory of
  Non-Equilibrium Systems}}}\ (\bibinfo  {publisher} {Cambridge University
  Press, Cambridge, UK},\ \bibinfo {year} {2011})\BibitemShut {NoStop}%
\bibitem [{\citenamefont {Meir}\ and\ \citenamefont
  {Wingreen}(1992)}]{PhysRevLett.68.2512}%
  \BibitemOpen
  \bibfield  {author} {\bibinfo {author} {\bibfnamefont {Y.}~\bibnamefont
  {Meir}}\ and\ \bibinfo {author} {\bibfnamefont {N.~S.}\ \bibnamefont
  {Wingreen}},\ }\href {\doibase 10.1103/PhysRevLett.68.2512} {\bibfield
  {journal} {\bibinfo  {journal} {Phys. Rev. Lett.}\ }\textbf {\bibinfo
  {volume} {68}},\ \bibinfo {pages} {2512} (\bibinfo {year}
  {1992})}\BibitemShut {NoStop}%
\bibitem [{\citenamefont {Fu}\ and\ \citenamefont
  {Kane}(2008)}]{FuKanePRL2008}%
  \BibitemOpen
  \bibfield  {author} {\bibinfo {author} {\bibfnamefont {L.}~\bibnamefont
  {Fu}}\ and\ \bibinfo {author} {\bibfnamefont {C.~L.}\ \bibnamefont {Kane}},\
  }\href {\doibase 10.1103/PhysRevLett.100.096407} {\bibfield  {journal}
  {\bibinfo  {journal} {Phys. Rev. Lett.}\ }\textbf {\bibinfo {volume} {100}},\
  \bibinfo {pages} {096407} (\bibinfo {year} {2008})}\BibitemShut {NoStop}%
\bibitem [{\citenamefont {He}\ \emph {et~al.}(2017)\citenamefont {He},
  \citenamefont {Pan}, \citenamefont {Stern}, \citenamefont {Burks},
  \citenamefont {Che}, \citenamefont {Yin}, \citenamefont {Wang}, \citenamefont
  {Lian}, \citenamefont {Zhou}, \citenamefont {Choi}, \citenamefont {Murata},
  \citenamefont {Kou}, \citenamefont {Chen}, \citenamefont {Nie}, \citenamefont
  {Shao}, \citenamefont {Fan}, \citenamefont {Zhang}, \citenamefont {Liu},
  \citenamefont {Xia},\ and\ \citenamefont {Wang}}]{HeScience2017}%
  \BibitemOpen
  \bibfield  {author} {\bibinfo {author} {\bibfnamefont {Q.~L.}\ \bibnamefont
  {He}}, \bibinfo {author} {\bibfnamefont {L.}~\bibnamefont {Pan}}, \bibinfo
  {author} {\bibfnamefont {A.~L.}\ \bibnamefont {Stern}}, \bibinfo {author}
  {\bibfnamefont {E.~C.}\ \bibnamefont {Burks}}, \bibinfo {author}
  {\bibfnamefont {X.}~\bibnamefont {Che}}, \bibinfo {author} {\bibfnamefont
  {G.}~\bibnamefont {Yin}}, \bibinfo {author} {\bibfnamefont {J.}~\bibnamefont
  {Wang}}, \bibinfo {author} {\bibfnamefont {B.}~\bibnamefont {Lian}}, \bibinfo
  {author} {\bibfnamefont {Q.}~\bibnamefont {Zhou}}, \bibinfo {author}
  {\bibfnamefont {E.~S.}\ \bibnamefont {Choi}}, \bibinfo {author}
  {\bibfnamefont {K.}~\bibnamefont {Murata}}, \bibinfo {author} {\bibfnamefont
  {X.}~\bibnamefont {Kou}}, \bibinfo {author} {\bibfnamefont {Z.}~\bibnamefont
  {Chen}}, \bibinfo {author} {\bibfnamefont {T.}~\bibnamefont {Nie}}, \bibinfo
  {author} {\bibfnamefont {Q.}~\bibnamefont {Shao}}, \bibinfo {author}
  {\bibfnamefont {Y.}~\bibnamefont {Fan}}, \bibinfo {author} {\bibfnamefont
  {S.-C.}\ \bibnamefont {Zhang}}, \bibinfo {author} {\bibfnamefont
  {K.}~\bibnamefont {Liu}}, \bibinfo {author} {\bibfnamefont {J.}~\bibnamefont
  {Xia}}, \ and\ \bibinfo {author} {\bibfnamefont {K.~L.}\ \bibnamefont
  {Wang}},\ }\href {\doibase 10.1126/science.aag2792} {\bibfield  {journal}
  {\bibinfo  {journal} {Science}\ }\textbf {\bibinfo {volume} {357}},\ \bibinfo
  {pages} {294} (\bibinfo {year} {2017})}\BibitemShut {NoStop}%
\bibitem [{\citenamefont {Shen}\ \emph {et~al.}(2020)\citenamefont {Shen},
  \citenamefont {Lyu}, \citenamefont {Gao}, \citenamefont {Xie}, \citenamefont
  {Chen}, \citenamefont {Cho}, \citenamefont {Atanov}, \citenamefont {Chen},
  \citenamefont {Liu}, \citenamefont {Hu}, \citenamefont {Yip}, \citenamefont
  {Goh}, \citenamefont {He}, \citenamefont {Pan}, \citenamefont {Wang},
  \citenamefont {Law},\ and\ \citenamefont {Lortz}}]{shen2018spectroscopic}%
  \BibitemOpen
  \bibfield  {author} {\bibinfo {author} {\bibfnamefont {J.}~\bibnamefont
  {Shen}}, \bibinfo {author} {\bibfnamefont {J.}~\bibnamefont {Lyu}}, \bibinfo
  {author} {\bibfnamefont {J.~Z.}\ \bibnamefont {Gao}}, \bibinfo {author}
  {\bibfnamefont {Y.-M.}\ \bibnamefont {Xie}}, \bibinfo {author} {\bibfnamefont
  {C.-Z.}\ \bibnamefont {Chen}}, \bibinfo {author} {\bibfnamefont {C.-w.}\
  \bibnamefont {Cho}}, \bibinfo {author} {\bibfnamefont {O.}~\bibnamefont
  {Atanov}}, \bibinfo {author} {\bibfnamefont {Z.}~\bibnamefont {Chen}},
  \bibinfo {author} {\bibfnamefont {K.}~\bibnamefont {Liu}}, \bibinfo {author}
  {\bibfnamefont {Y.~J.}\ \bibnamefont {Hu}}, \bibinfo {author} {\bibfnamefont
  {K.~Y.}\ \bibnamefont {Yip}}, \bibinfo {author} {\bibfnamefont {S.~K.}\
  \bibnamefont {Goh}}, \bibinfo {author} {\bibfnamefont {Q.~L.}\ \bibnamefont
  {He}}, \bibinfo {author} {\bibfnamefont {L.}~\bibnamefont {Pan}}, \bibinfo
  {author} {\bibfnamefont {K.~L.}\ \bibnamefont {Wang}}, \bibinfo {author}
  {\bibfnamefont {K.~T.}\ \bibnamefont {Law}}, \ and\ \bibinfo {author}
  {\bibfnamefont {R.}~\bibnamefont {Lortz}},\ }\href {\doibase
  10.1073/pnas.1910967117} {\bibfield  {journal} {\bibinfo  {journal}
  {Proceedings of the National Academy of Sciences}\ }\textbf {\bibinfo
  {volume} {117}},\ \bibinfo {pages} {238} (\bibinfo {year}
  {2020})}\BibitemShut {NoStop}%
\bibitem [{Note1()}]{Note1}%
  \BibitemOpen
  \bibinfo {note} {The sums $\DOTSB \sum@ \slimits@ _{n}\protect \frac
  {\protect \qopname \relax o{cos}A n}{n^2}$ or $\DOTSB \sum@ \slimits@
  _{n}\protect \frac {\protect \qopname \relax o{sin}A n}{n^2}$ are reduced to
  the polylogarithm function ${\protect \rm Li}_2(z){=}\DOTSB \sum@ \slimits@
  _{k=1}^\infty \protect \frac {z^k}{k^2}$ which has the property ${\protect
  \rm Re}[{\protect \rm Li}_2(e^{2\pi i z}) ]{=}\pi ^2 ({\protect \rm mod}_1(z)
  {-}\protect \frac {1}{2})^2{-}\protect \frac {\pi ^2}{12}$ for real
  $z$.}\BibitemShut {Stop}%
\bibitem [{\citenamefont {Fu}\ and\ \citenamefont
  {Kane}(2009)}]{FuKanePRL2009}%
  \BibitemOpen
  \bibfield  {author} {\bibinfo {author} {\bibfnamefont {L.}~\bibnamefont
  {Fu}}\ and\ \bibinfo {author} {\bibfnamefont {C.~L.}\ \bibnamefont {Kane}},\
  }\href {\doibase 10.1103/PhysRevLett.102.216403} {\bibfield  {journal}
  {\bibinfo  {journal} {Phys. Rev. Lett.}\ }\textbf {\bibinfo {volume} {102}},\
  \bibinfo {pages} {216403} (\bibinfo {year} {2009})}\BibitemShut {NoStop}%
\bibitem [{\citenamefont {Akhmerov}\ \emph {et~al.}(2009)\citenamefont
  {Akhmerov}, \citenamefont {Nilsson},\ and\ \citenamefont
  {Beenakker}}]{PhysRevLett.102.216404}%
  \BibitemOpen
  \bibfield  {author} {\bibinfo {author} {\bibfnamefont {A.~R.}\ \bibnamefont
  {Akhmerov}}, \bibinfo {author} {\bibfnamefont {J.}~\bibnamefont {Nilsson}}, \
  and\ \bibinfo {author} {\bibfnamefont {C.~W.~J.}\ \bibnamefont {Beenakker}},\
  }\href {\doibase 10.1103/PhysRevLett.102.216404} {\bibfield  {journal}
  {\bibinfo  {journal} {Phys. Rev. Lett.}\ }\textbf {\bibinfo {volume} {102}},\
  \bibinfo {pages} {216404} (\bibinfo {year} {2009})}\BibitemShut {NoStop}%
\bibitem [{\citenamefont {Str\"ubi}\ \emph {et~al.}(2015)\citenamefont
  {Str\"ubi}, \citenamefont {Belzig}, \citenamefont {Schmidt},\ and\
  \citenamefont {Bruder}}]{STRUBI2015489}%
  \BibitemOpen
  \bibfield  {author} {\bibinfo {author} {\bibfnamefont {G.}~\bibnamefont
  {Str\"ubi}}, \bibinfo {author} {\bibfnamefont {W.}~\bibnamefont {Belzig}},
  \bibinfo {author} {\bibfnamefont {T.~L.}\ \bibnamefont {Schmidt}}, \ and\
  \bibinfo {author} {\bibfnamefont {C.}~\bibnamefont {Bruder}},\ }\href
  {\doibase https://doi.org/10.1016/j.physe.2015.08.005} {\bibfield  {journal}
  {\bibinfo  {journal} {Physica E: Low-dimensional Systems and Nanostructures}\
  }\textbf {\bibinfo {volume} {74}},\ \bibinfo {pages} {489 } (\bibinfo {year}
  {2015})}\BibitemShut {NoStop}%
\bibitem [{\citenamefont {Str\"ubi}\ \emph {et~al.}(2011)\citenamefont
  {Str\"ubi}, \citenamefont {Belzig}, \citenamefont {Choi},\ and\ \citenamefont
  {Bruder}}]{PhysRevLett.107.136403}%
  \BibitemOpen
  \bibfield  {author} {\bibinfo {author} {\bibfnamefont {G.}~\bibnamefont
  {Str\"ubi}}, \bibinfo {author} {\bibfnamefont {W.}~\bibnamefont {Belzig}},
  \bibinfo {author} {\bibfnamefont {M.-S.}\ \bibnamefont {Choi}}, \ and\
  \bibinfo {author} {\bibfnamefont {C.}~\bibnamefont {Bruder}},\ }\href
  {\doibase 10.1103/PhysRevLett.107.136403} {\bibfield  {journal} {\bibinfo
  {journal} {Phys. Rev. Lett.}\ }\textbf {\bibinfo {volume} {107}},\ \bibinfo
  {pages} {136403} (\bibinfo {year} {2011})}\BibitemShut {NoStop}%
\bibitem [{\citenamefont {Li}\ \emph {et~al.}(2012)\citenamefont {Li},
  \citenamefont {Fleury},\ and\ \citenamefont
  {B\"uttiker}}]{PhysRevB.85.125440}%
  \BibitemOpen
  \bibfield  {author} {\bibinfo {author} {\bibfnamefont {J.}~\bibnamefont
  {Li}}, \bibinfo {author} {\bibfnamefont {G.}~\bibnamefont {Fleury}}, \ and\
  \bibinfo {author} {\bibfnamefont {M.}~\bibnamefont {B\"uttiker}},\ }\href
  {\doibase 10.1103/PhysRevB.85.125440} {\bibfield  {journal} {\bibinfo
  {journal} {Phys. Rev. B}\ }\textbf {\bibinfo {volume} {85}},\ \bibinfo
  {pages} {125440} (\bibinfo {year} {2012})}\BibitemShut {NoStop}%
\bibitem [{\citenamefont {Shapiro}\ \emph {et~al.}(2016)\citenamefont
  {Shapiro}, \citenamefont {Shnirman},\ and\ \citenamefont
  {Mirlin}}]{PhysRevB.93.155411}%
  \BibitemOpen
  \bibfield  {author} {\bibinfo {author} {\bibfnamefont {D.~S.}\ \bibnamefont
  {Shapiro}}, \bibinfo {author} {\bibfnamefont {A.}~\bibnamefont {Shnirman}}, \
  and\ \bibinfo {author} {\bibfnamefont {A.~D.}\ \bibnamefont {Mirlin}},\
  }\href {\doibase 10.1103/PhysRevB.93.155411} {\bibfield  {journal} {\bibinfo
  {journal} {Phys. Rev. B}\ }\textbf {\bibinfo {volume} {93}},\ \bibinfo
  {pages} {155411} (\bibinfo {year} {2016})}\BibitemShut {NoStop}%
\bibitem [{\citenamefont {Shapiro}\ \emph {et~al.}(2017)\citenamefont
  {Shapiro}, \citenamefont {Feldman}, \citenamefont {Mirlin},\ and\
  \citenamefont {Shnirman}}]{PhysRevB.95.195425}%
  \BibitemOpen
  \bibfield  {author} {\bibinfo {author} {\bibfnamefont {D.~S.}\ \bibnamefont
  {Shapiro}}, \bibinfo {author} {\bibfnamefont {D.~E.}\ \bibnamefont
  {Feldman}}, \bibinfo {author} {\bibfnamefont {A.~D.}\ \bibnamefont {Mirlin}},
  \ and\ \bibinfo {author} {\bibfnamefont {A.}~\bibnamefont {Shnirman}},\
  }\href {\doibase 10.1103/PhysRevB.95.195425} {\bibfield  {journal} {\bibinfo
  {journal} {Phys. Rev. B}\ }\textbf {\bibinfo {volume} {95}},\ \bibinfo
  {pages} {195425} (\bibinfo {year} {2017})}\BibitemShut {NoStop}%
\bibitem [{\citenamefont {Shapiro}\ \emph {et~al.}(2018)\citenamefont
  {Shapiro}, \citenamefont {Mirlin},\ and\ \citenamefont
  {Shnirman}}]{PhysRevB.98.245405}%
  \BibitemOpen
  \bibfield  {author} {\bibinfo {author} {\bibfnamefont {D.~S.}\ \bibnamefont
  {Shapiro}}, \bibinfo {author} {\bibfnamefont {A.~D.}\ \bibnamefont {Mirlin}},
  \ and\ \bibinfo {author} {\bibfnamefont {A.}~\bibnamefont {Shnirman}},\
  }\href {\doibase 10.1103/PhysRevB.98.245405} {\bibfield  {journal} {\bibinfo
  {journal} {Phys. Rev. B}\ }\textbf {\bibinfo {volume} {98}},\ \bibinfo
  {pages} {245405} (\bibinfo {year} {2018})}\BibitemShut {NoStop}%
\bibitem [{\citenamefont {Chung}\ \emph {et~al.}(2011)\citenamefont {Chung},
  \citenamefont {Qi}, \citenamefont {Maciejko},\ and\ \citenamefont
  {Zhang}}]{PhysRevB.83.100512}%
  \BibitemOpen
  \bibfield  {author} {\bibinfo {author} {\bibfnamefont {S.~B.}\ \bibnamefont
  {Chung}}, \bibinfo {author} {\bibfnamefont {X.-L.}\ \bibnamefont {Qi}},
  \bibinfo {author} {\bibfnamefont {J.}~\bibnamefont {Maciejko}}, \ and\
  \bibinfo {author} {\bibfnamefont {S.-C.}\ \bibnamefont {Zhang}},\ }\href
  {\doibase 10.1103/PhysRevB.83.100512} {\bibfield  {journal} {\bibinfo
  {journal} {Phys. Rev. B}\ }\textbf {\bibinfo {volume} {83}},\ \bibinfo
  {pages} {100512(R)} (\bibinfo {year} {2011})}\BibitemShut {NoStop}%
\bibitem [{\citenamefont {Liu}\ and\ \citenamefont
  {Trauzettel}(2011)}]{PhysRevB.83.220510}%
  \BibitemOpen
  \bibfield  {author} {\bibinfo {author} {\bibfnamefont {C.-X.}\ \bibnamefont
  {Liu}}\ and\ \bibinfo {author} {\bibfnamefont {B.}~\bibnamefont
  {Trauzettel}},\ }\href {\doibase 10.1103/PhysRevB.83.220510} {\bibfield
  {journal} {\bibinfo  {journal} {Phys. Rev. B}\ }\textbf {\bibinfo {volume}
  {83}},\ \bibinfo {pages} {220510(R)} (\bibinfo {year} {2011})}\BibitemShut
  {NoStop}%
\bibitem [{\citenamefont {Hou}\ \emph {et~al.}(2013)\citenamefont {Hou},
  \citenamefont {Shtengel},\ and\ \citenamefont {Refael}}]{PhysRevB.88.075304}%
  \BibitemOpen
  \bibfield  {author} {\bibinfo {author} {\bibfnamefont {C.-Y.}\ \bibnamefont
  {Hou}}, \bibinfo {author} {\bibfnamefont {K.}~\bibnamefont {Shtengel}}, \
  and\ \bibinfo {author} {\bibfnamefont {G.}~\bibnamefont {Refael}},\ }\href
  {\doibase 10.1103/PhysRevB.88.075304} {\bibfield  {journal} {\bibinfo
  {journal} {Phys. Rev. B}\ }\textbf {\bibinfo {volume} {88}},\ \bibinfo
  {pages} {075304} (\bibinfo {year} {2013})}\BibitemShut {NoStop}%
\bibitem [{\citenamefont {Lian}\ \emph {et~al.}(2018)\citenamefont {Lian},
  \citenamefont {Sun}, \citenamefont {Vaezi}, \citenamefont {Qi},\ and\
  \citenamefont {Zhang}}]{Lian10938}%
  \BibitemOpen
  \bibfield  {author} {\bibinfo {author} {\bibfnamefont {B.}~\bibnamefont
  {Lian}}, \bibinfo {author} {\bibfnamefont {X.-Q.}\ \bibnamefont {Sun}},
  \bibinfo {author} {\bibfnamefont {A.}~\bibnamefont {Vaezi}}, \bibinfo
  {author} {\bibfnamefont {X.-L.}\ \bibnamefont {Qi}}, \ and\ \bibinfo {author}
  {\bibfnamefont {S.-C.}\ \bibnamefont {Zhang}},\ }\href {\doibase
  10.1073/pnas.1810003115} {\bibfield  {journal} {\bibinfo  {journal}
  {Proceedings of the National Academy of Sciences}\ }\textbf {\bibinfo
  {volume} {115}},\ \bibinfo {pages} {10938} (\bibinfo {year}
  {2018})}\BibitemShut {NoStop}%
\bibitem [{\citenamefont {Beenakker}\ \emph {et~al.}(2019)\citenamefont
  {Beenakker}, \citenamefont {Baireuther}, \citenamefont {Herasymenko},
  \citenamefont {Adagideli}, \citenamefont {Wang},\ and\ \citenamefont
  {Akhmerov}}]{PhysRevLett.122.146803}%
  \BibitemOpen
  \bibfield  {author} {\bibinfo {author} {\bibfnamefont {C.~W.~J.}\
  \bibnamefont {Beenakker}}, \bibinfo {author} {\bibfnamefont {P.}~\bibnamefont
  {Baireuther}}, \bibinfo {author} {\bibfnamefont {Y.}~\bibnamefont
  {Herasymenko}}, \bibinfo {author} {\bibfnamefont {I.}~\bibnamefont
  {Adagideli}}, \bibinfo {author} {\bibfnamefont {L.}~\bibnamefont {Wang}}, \
  and\ \bibinfo {author} {\bibfnamefont {A.~R.}\ \bibnamefont {Akhmerov}},\
  }\href {\doibase 10.1103/PhysRevLett.122.146803} {\bibfield  {journal}
  {\bibinfo  {journal} {Phys. Rev. Lett.}\ }\textbf {\bibinfo {volume} {122}},\
  \bibinfo {pages} {146803} (\bibinfo {year} {2019})}\BibitemShut {NoStop}%
\bibitem [{\citenamefont {Hassler}\ \emph {et~al.}(2020)\citenamefont
  {Hassler}, \citenamefont {Grabsch}, \citenamefont {Pacholski}, \citenamefont
  {Oriekhov}, \citenamefont {Ovdat}, \citenamefont {Adagideli},\ and\
  \citenamefont {Beenakker}}]{PhysRevB.102.045431}%
  \BibitemOpen
  \bibfield  {author} {\bibinfo {author} {\bibfnamefont {F.}~\bibnamefont
  {Hassler}}, \bibinfo {author} {\bibfnamefont {A.}~\bibnamefont {Grabsch}},
  \bibinfo {author} {\bibfnamefont {M.~J.}\ \bibnamefont {Pacholski}}, \bibinfo
  {author} {\bibfnamefont {D.~O.}\ \bibnamefont {Oriekhov}}, \bibinfo {author}
  {\bibfnamefont {O.}~\bibnamefont {Ovdat}}, \bibinfo {author} {\bibfnamefont
  {I.}~\bibnamefont {Adagideli}}, \ and\ \bibinfo {author} {\bibfnamefont
  {C.~W.~J.}\ \bibnamefont {Beenakker}},\ }\href {\doibase
  10.1103/PhysRevB.102.045431} {\bibfield  {journal} {\bibinfo  {journal}
  {Phys. Rev. B}\ }\textbf {\bibinfo {volume} {102}},\ \bibinfo {pages}
  {045431} (\bibinfo {year} {2020})}\BibitemShut {NoStop}%
\bibitem [{\citenamefont {Beenakker}\ and\ \citenamefont
  {Oriekhov}(2020)}]{beenakker2020shot}%
  \BibitemOpen
  \bibfield  {author} {\bibinfo {author} {\bibfnamefont {C.}~\bibnamefont
  {Beenakker}}\ and\ \bibinfo {author} {\bibfnamefont {D.}~\bibnamefont
  {Oriekhov}},\ }\href@noop {} {\bibfield  {journal} {\bibinfo  {journal}
  {SciPost Physics}\ }\textbf {\bibinfo {volume} {9}},\ \bibinfo {pages} {080}
  (\bibinfo {year} {2020})}\BibitemShut {NoStop}%
\bibitem [{\citenamefont {Adagideli}\ \emph {et~al.}(2020)\citenamefont
  {Adagideli}, \citenamefont {Hassler}, \citenamefont {Grabsch}, \citenamefont
  {Pacholski},\ and\ \citenamefont {Beenakker}}]{AdagideliSciPost2019}%
  \BibitemOpen
  \bibfield  {author} {\bibinfo {author} {\bibfnamefont {I.}~\bibnamefont
  {Adagideli}}, \bibinfo {author} {\bibfnamefont {F.}~\bibnamefont {Hassler}},
  \bibinfo {author} {\bibfnamefont {A.}~\bibnamefont {Grabsch}}, \bibinfo
  {author} {\bibfnamefont {M.}~\bibnamefont {Pacholski}}, \ and\ \bibinfo
  {author} {\bibfnamefont {C.}~\bibnamefont {Beenakker}},\ }\href {\doibase
  10.21468/SciPostPhys.8.1.013} {\bibfield  {journal} {\bibinfo  {journal}
  {SciPost Phys.}\ }\textbf {\bibinfo {volume} {8}},\ \bibinfo {pages} {13}
  (\bibinfo {year} {2020})}\BibitemShut {NoStop}%
\bibitem [{\citenamefont {Shapiro}\ \emph {et~al.}(2021)\citenamefont
  {Shapiro}, \citenamefont {Mirlin},\ and\ \citenamefont
  {Shnirman}}]{PhysRevB.104.035434}%
  \BibitemOpen
  \bibfield  {author} {\bibinfo {author} {\bibfnamefont {D.~S.}\ \bibnamefont
  {Shapiro}}, \bibinfo {author} {\bibfnamefont {A.~D.}\ \bibnamefont {Mirlin}},
  \ and\ \bibinfo {author} {\bibfnamefont {A.}~\bibnamefont {Shnirman}},\
  }\href {\doibase 10.1103/PhysRevB.104.035434} {\bibfield  {journal} {\bibinfo
   {journal} {Phys. Rev. B}\ }\textbf {\bibinfo {volume} {104}},\ \bibinfo
  {pages} {035434} (\bibinfo {year} {2021})}\BibitemShut {NoStop}%
\bibitem [{\citenamefont {Altland}\ and\ \citenamefont
  {Simons}(2010)}]{altland2010condensed}%
  \BibitemOpen
  \bibfield  {author} {\bibinfo {author} {\bibfnamefont {A.}~\bibnamefont
  {Altland}}\ and\ \bibinfo {author} {\bibfnamefont {B.~D.}\ \bibnamefont
  {Simons}},\ }\href@noop {} {\emph {\bibinfo {title} {Condensed matter field
  theory}}}\ (\bibinfo  {publisher} {Cambridge university press},\ \bibinfo
  {year} {2010})\BibitemShut {NoStop}%
\end{thebibliography}
\end{document}